\documentclass[letterpaper,twocolumn,pra,aps,showpacs,groupaddress,
floatfix]{revtex4-1}
\usepackage[utf8]{inputenc}
\usepackage{bm}
\usepackage[usenames]{color}
\usepackage{multirow}
\usepackage{amssymb}
\usepackage{amsbsy}
\usepackage{amsmath}
\usepackage{stmaryrd}
\usepackage{graphicx}
\usepackage{epsfig}
\usepackage{placeins}
\usepackage[normalem]{ulem}
\usepackage{bbold}
\usepackage{bbm}
\usepackage{braket}
\usepackage{graphicx}
\usepackage{tikz}
\usepackage[colorlinks,linkcolor=blue,citecolor=blue,urlcolor=blue]{hyperref}

\makeatletter

\setcitestyle{numbers,square,comma,sort&compress}

\newcommand{\figref}[1]{Fig.~\protect\ref{#1}}

\newcommand{\subeq} {_{\mathrm{eq}}}

\newcommand{\Tr}[1] {\mathrm{Tr}\left[\,{#1}\,\right]}
\newcommand{\intpic} {_{\mathrm{I}}}
\newcommand{\emath}[1]{\mathrm{e}^{#1}}
\newcommand{\timevolve}[2]{\emath{-#2}\,{#1}\,\emath{#2}}
\newcommand{\commutator}[2]{\left[\,#1,#2\,\right]}

\begin{document}
\title{Selected applications of typicality to real-time dynamics of quantum 
many-body systems}

\author{Tjark Heitmann}
\email{tjark.heitmann@uos.de}
\affiliation{Department of Physics, University of Osnabr\"uck, D-49069 
Osnabr\"uck, Germany}

\author{Jonas Richter}
\affiliation{Department of Physics, University of Osnabr\"uck, D-49069 
Osnabr\"uck, Germany}

\author{Dennis Schubert}
\affiliation{Department of Physics, University of Osnabr\"uck, D-49069 
Osnabr\"uck, Germany}

\author{Robin Steinigeweg}
\email{rsteinig@uos.de}
\affiliation{Department of Physics, University of Osnabr\"uck, D-49069 
Osnabr\"uck, Germany}

\date{\today}

\begin{abstract}

Loosely speaking, the concept of quantum typicality refers to the fact that a 
single pure state can imitate the full statistical ensemble. This fact 
has given rise to a rather simple 
but remarkably useful numerical approach to simulate the dynamics of quantum 
many-body systems, called {\it dynamical quantum 
typicality} (DQT). In this paper, we give a 
brief overview of selected applications of DQT, where particular emphasis 
is given to questions on transport and thermalization in low-dimensional 
lattice systems like chains or ladders of interacting spins or fermions. For these 
systems, we discuss 
that DQT provides an efficient means to obtain time-dependent equilibrium 
correlation functions for comparatively large Hilbert-space dimensions and long 
time scales, allowing the quantitative extraction of transport coefficients 
within the framework of, e.g., linear response theory. Furthermore, 
it is discussed that DQT can also be used to study the far-from-equilibrium 
dynamics resulting from 
sudden quench scenarios, where the initial state is a thermal Gibbs state 
of the pre-quench Hamiltonian. Eventually, we summarize a few combinations 
of DQT with other approaches such as numerical linked cluster 
expansions or 
projection operator techniques. In this way, we demonstrate the versatility of DQT. 

\end{abstract}

\maketitle


\section{Introduction}

Unraveling the dynamics of isolated quantum many-body systems 
is a central objective of modern experimental and theoretical physics. On the 
one hand, new experimental platforms composed of cold atoms or trapped ions 
have opened the door to perform quantum simulations with a high amount 
of control over Hamiltonian parameters and initial conditions 
\cite{Bloch2012, Blatt2012}. On the other hand,
there has been substantial progress from the theoretical side to understand (i) 
experimental observations and (ii) long-standing 
questions about the fundamentals of statistical mechanics 
\cite{Polkovnikov2011, Eisert2015, Dalessio2016, Gogolin2016, Borgonovi2016}. 
One such question is how to reconcile the emergence of thermodynamic behavior 
with the unitary time 
evolution of isolated quantum systems, i.e., to explain whether and in which way an 
isolated system relaxes towards a stationary long-time state 
which agrees with the predictions from standard statistical mechanics.
Another similarly intriguing question in this context is to explain the onset of 
conventional hydrodynamic transport, i.e., diffusion, from truly microscopic 
principles \cite{Buchanan2005, Khemani2018, Bertini2020}. The numerical analysis of 
thermalization and transport in isolated 
quantum many-body systems is at the heart of this paper.  

Generally, the theoretical analysis of quantum 
many-body dynamics is notoriously difficult. Given a quantum system ${\cal H}$ 
and an arbitrary nonequilibrium state $\rho(0)$, universal concepts to 
describe the resulting dynamics are rare \cite{Pruefer2018, 
Erne2018, Richter2019c}, and one is usually required to 
solve the microscopic equation of motion for the density matrix 
$\rho(t)$, i.e., the von-Neumann equation   
\begin{equation}\label{Eq::vonNeumann}
 \frac{\text{d}}{\text{d}t} \rho(t) = -i[{\cal H},\rho(t)]
\end{equation}
($\hbar=1$) which, in the case of a pure state $\rho(t) = 
\ket{\psi(t)}\bra{\psi(t)}$, reduces to the Schr\"odinger 
equation
\begin{equation}\label{Eq::schroedi}
	\frac{\text{d}}{\text{d}t} \ket{\psi(t)} = -i{\cal H}\ket{\psi(t)}\ . 
\end{equation}
While the presence of strong interactions often prohibits any 
analytical solution, numerical studies of Eq.\ \eqref{Eq::schroedi} are 
plagued by the exponential growth of the Hilbert space upon 
increasing the number of degrees of freedom. Moreover, since thermalization and 
transport can potentially be very slow processes, the necessity to study 
long time scales adds 
another layer of complexity. 

Of course, for situations close to equilibrium, 
e.g., a system being weakly perturbed by an external force, linear response 
theory provides a successful framework to 
describe the system's response in terms of dynamical correlation functions 
evaluated exactly at equilibrium \cite{Kubo1991}. However, analogous to Eqs.\ 
\eqref{Eq::vonNeumann} and \eqref{Eq::schroedi}, the calculation of 
such time-dependent correlation 
functions for large system sizes and long time scales is a severe challenge in 
practice. 

Despite these difficulties, significant progress has been made over the 
years thanks to the augmented availability of computational resources and 
the development of sophisticated numerical techniques. 
Especially 
for one-dimensional systems the time-dependent density matrix renormalization 
group (tDMRG), including related methods based on matrix product states, 
provides a 
powerful approach to dynamical properties in the thermodynamic 
limit (for reviews, see \cite{Schollwoeck2011, Paeckel2019}). 
However, due to the inevitable 
build-up of entanglement, this approach is limited in the
time scales which can be reached in simulations. 

In the present paper, the focus is on another
useful numerical approach to the dynamics of quantum many-body systems, 
which is based on the concept of dynamical quantum typicality (DQT) 
\cite{Bartsch2009, Reimann2018}. 
In a nutshell, DQT means that ``the vast majority of all pure states featuring a common expectation value of some generic
observable at a given time will yield very similar expectation values of 
the same observable at any later time'' \cite{Bartsch2009}.
In fact, the idea of using 
random vectors has a long and fruitful history \cite{Alben1975,Lloyd1988,
DeRaedt1989, Jaklic1994, Gemmer2004, Popescu2006, Goldstein2006, Reimann2007}. 
By virtue of an iterative forward propagation of 
these vectors in real or imaginary time, dependencies on time and temperature can 
be obtained. Since DQT can be implemented rather 
memory efficiently, it is possible to study dynamical properties of quantum 
many-body systems with Hilbert-space dimensions significantly larger compared 
to standard exact diagonalization (ED). Moreover, there are no conceptual 
limitations on the reachable time scales.     

It is worth pointing out that DQT can not only be used to obtain time-dependent 
properties \cite{Iitaka2003, Elsayed2013, Steinigeweg2014} or spectral functions
\cite{Jaklic1994, Okamoto2018, Yamaji2018, Rousochatzakis2019}
but also static properties such as the density of states 
\cite{Hams2000} or thermodynamic quantities 
\cite{DeVries1993, Sugiura2012, Sugiura2013, Wietek2019}. 
However, it is the aim of this paper to discuss the 
usefulness and versatility of DQT especially in the context of thermalization 
and transport. 

This paper is structured as follows. In Sec.\ \ref{Sec::Typ}, 
we give a brief introduction to the concept of typicality and 
also elaborate on the differences between typicality and the eigenstate 
thermalization hypothesis. In 
Sec.\ \ref{Sec::Num}, we discuss various applications of typicality to the 
dynamics of quantum many-body systems. Finally, we summarize and 
conclude in Sec.\ \ref{Sec::Sum}, where we also provide an outlook on 
further applications of DQT. 

\section{What is typicality?}\label{Sec::Typ}
 
Loosely speaking, the notion of typicality means that even a 
single pure quantum state can imitate the full statistical ensemble, or, more precisely, expectation values of typical pure states are close to the expectation value of the statistical ensemble
\cite{Lloyd1988, Reimann2007, Popescu2006, Goldstein2006, Gemmer2004}.
While typicality has been put forward as an important insight to explain the 
emergence of thermodynamic behavior (see e.g.\ Ref.\ \cite{Gemmer2004} for 
an overview), let us here focus on the practical consequences of typicality. 
In particular, let us consider the, e.g., canonical equilibrium expectation 
value $\langle A \rangle_\text{eq}$ of some (quasi-local) operator $A$ defined as 
\begin{equation}\label{Eq::Equilibrium}
 \langle A \rangle_\text{eq} = \frac{\text{Tr}[A e^{-\beta {\cal 
H}}]}{\mathcal{Z}} = \frac{\text{Tr}[e^{-\beta {\cal 
H}/2} A e^{-\beta {\cal 
H}/2}]}{\mathcal{Z}}\ , 
\end{equation}
where $\mathcal{Z}=\text{Tr}\left[\exp(-\beta\mathcal{H})\right]$ is the 
canonical partition
function,
$\beta = 1/T$ ($k_\text{B}=1$) is the inverse temperature, and we have used the cyclic 
invariance of the trace. Exploiting 
typicality, it is possible to rewrite $\langle A \rangle_\text{eq}$ according 
to
\begin{equation} \label{Eq::DQTApprox}
 \langle A \rangle_\text{eq} = 
\frac{\bra{\psi_\beta}A\ket{\psi_\beta}}{\braket{\psi_\beta|\psi_\beta}} + 
\epsilon\ , 
\end{equation}
where we have introduced the abbreviation $\ket{\psi_\beta} = e^{-\beta {\cal 
H}/2}\ket{\psi}$, which is sometimes referred to as thermal pure quantum state 
\cite{Sugiura2013}. The reference pure 
state $\ket{\psi}$ is drawn at random from the full Hilbert space with finite
dimension $d$ according to the unitary invariant Haar measure 
\cite{Bartsch2009}, i.e., 
\begin{equation} \label{Eq::RandState}
 \ket{\psi} = \sum_{k=1}^d (a_k + i b_k) \ket{k}\ , 
\end{equation}
where the coefficients $a_k$ and $b_k$ are drawn from a Gaussian distribution 
with zero 
mean (other types of randomness have been suggested 
as well \cite{Alben1975,Iitaka2004}), and the pure states 
$\ket{k}$ denote orthogonal basis states of the Hilbert space. 
(Note that $\ket{\psi}$ is almost maximally entangled 
\cite{Page1993,Vidmar2017}.) 
Importantly, the variance of the statistical error $\epsilon = \epsilon(\ket{\psi})$ of the 
approximation \eqref{Eq::DQTApprox} 
scales as $\sigma \propto 1/\sqrt{d_\text{eff}}$, where $d_\text{eff} = 
\text{Tr}[\exp(-\beta({\cal H}-E_0))]$ is the effective dimension of the Hilbert 
space with $E_0$ being the ground-state energy of ${\cal H}$. Here we assume that $A$ is a local operator (or a low-degree polynomial in system size), which applies to all examples discussed in this paper.
For more details on  error bounds see, e.g., Refs.\ \cite{Sugiura2013,Reimann2018}. For empirical estimates, see, e.g., Ref.\ \cite{Schnack2020}.
Thus, $d_\text{eff}$ 
is essentially the number of thermally occupied states and, for $\beta = 0$, we 
have $d_\text{eff} = d$. As a consequence, increasing the number of 
degrees of freedom of a quantum many-body system, e.g., the number of 
lattice sites $L$, leads to an exponential improvement of the  
accuracy (the higher the temperature, the faster), and Eq.\ 
\eqref{Eq::DQTApprox} becomes 
exact in the thermodynamic limit $L \to \infty$. 

The typicality approximation \eqref{Eq::DQTApprox} has 
proven to be very useful to calculate equilibrium quantities of quantum 
many-body systems such as the specific heat, entropy, or magnetic 
susceptibility \cite{Sugiura2012, Sugiura2013, DeVries1993, 
Wietek2019, Schnack2020}. 
For the purpose of this 
review, however, it is most
important to note that typicality is not just restricted to equilibrium 
properties, but also extends to the real-time dynamics of quantum expectation 
values \cite{Iitaka2003, Bartsch2009, Elsayed2013, Monnai2014, Steinigeweg2014, Endo2018, Richter2019d}. This dynamical version of typicality forms the 
basis of 
the numerical approach to time-dependent correlation functions and 
out-of-equilibrium dynamics more generally, which is 
discussed in Sec.\ \ref{Sec::Num}.

Let us briefly discuss the relationship between typicality and 
the eigenstate thermalization hypothesis (ETH) \cite{Deutsch1991, 
Srednicki1994, Rigol2008}. The ETH 
states that the expectation values of local observables 
evaluated within individual eigenstates $\ket{n}$ of generic (nonintegrable) Hamiltonians 
coincide with the microcanonical ensemble average at the 
corresponding energy density,
\begin{equation}
 A_{nn} = \bra{n} A \ket{n} = A_\text{mc}(E)\ . 
\end{equation}
While this fact (i.e.\ pure states can 
approximate ensemble expectation values) appears similar to our discussion of 
typicality in the 
context of Eq.\ \eqref{Eq::DQTApprox}, let us stress that typicality and ETH 
are two 
distinct concepts. On the one hand, while the ETH is assumed to hold for a 
variety of few-body operators and nonintegrable 
models \cite{Santos2010, Steinigeweg2013, Beugeling2014, Kim2014, Nandkishore2015, 
Dalessio2016, Mondaini2017, Jansen2019, Khaymovich2019}, a 
rigorous proof for its validity is still absent. 
On the other hand, 
typicality is no assumption and essentially requires the largeness of 
the effective Hilbert-space dimension. This difference becomes particularly
clear from the following point of view: since the 
distribution of the $a_k$ and $b_k$ in Eq.\ \eqref{Eq::RandState} is invariant 
under any unitary transformation, the state $\ket{\psi}$ is a random 
superposition also in the eigenbasis of ${\cal H}$ (whereas the ETH 
just refers to single eigenstates). Due to this randomness, Eq.\ 
\eqref{Eq::DQTApprox} holds even in cases where the ETH breaks 
down, i.e., where the expectation values of observables 
exhibit strong eigenstate-to-eigenstate fluctuations.

Since typicality is 
independent of the validity of the ETH, it can be used in integrable 
or many-body localized models, where 
the ETH is expected to be violated \cite{Steinigeweg2016a, Steinigeweg2017a,  
Richter2018, Richter2019g}. As a side remark, typicality can also be used to
test the ETH \cite{Steinigeweg2014a}. 

Eventually, let us emphasize that 
the choice of the specific basis $\ket{k}$ in Eq.\ \eqref{Eq::RandState} is 
arbitrary. Therefore, the random state $\ket{\psi}$ can be conveniently 
constructed in the working basis which is used to set up the Hamiltonian and 
all other observables. For instance, when working with spin-$1/2$ systems, a 
common choice is the so-called Ising basis, i.e., the states 
$\ket{k}$ then denote the $2^L$ different combinations of $\uparrow$ and 
$\downarrow$. Naturally, it is 
possible to combine DQT with the use of symmetries \cite{Heitmann2019}, where 
a random state is then drawn independently within each subsector.   

\section{DQT as a numerical tool}\label{Sec::Num}

We now discuss the use of dynamical 
quantum typicality as a numerical method. To begin with, we discuss 
in Sec.\ \ref{Sec::PSP} the iterative forward propagation of pure states 
in large Hilbert spaces. Afterwards, as a first application, we demonstrate in 
Sec.\ \ref{Sec::DOS} how typicality can be used to study the (local) density 
of states. In Sec.\ \ref{Sec::Trans}, we then show how DQT 
can be used to evaluate equilibrium correlation functions within the framework of 
linear response theory. Sec.\ \ref{Sec::FFED} is concerned with the 
out-of-equilibrium dynamics in certain quantum-quench scenarios. Eventually, in 
Sec.\ \ref{Sec::Ext}, we 
discuss how DQT can be combined with other approaches such as numerical 
linked cluster expansions or projection operator techniques.  

\subsection{Pure-state propagation}\label{Sec::PSP}

From a numerical point of view, a central advantage of the typicality approach 
comes from the fact that one can work with pure states instead of having to deal 
with full density matrices. This fact leads to a substantial reduction of the 
memory requirements, since it is possible to efficiently generate time and 
temperature dependencies of pure states. (Note that, while it is always possible to purify a density matrix, the DQT approach in contrast does not require to square the Hilbert-space dimension \cite{Hughston1993}.)

Specifically, let us consider the pure state $\ket{\psi_\beta} = e^{-\beta 
{\cal H}/2}\ket{\psi}$ introduced in Eq.\ \eqref{Eq::DQTApprox}. 
The time evolution of $\ket{\psi_\beta}$ is given by 
$\ket{\psi_\beta(t)} = e^{-i{\cal H}t}\ket{\psi_\beta}$. 
The 
full evolution up to time $t$ can be subdivided into a product of consecutive 
steps,
\begin{equation}
 \ket{\psi_\beta(t)} = \left(e^{-i{\cal H}\delta t}\right)^N \ket{\psi_\beta}\ 
, 
\end{equation}
where $\delta t = t/N$ is a discrete time step. 
If $\delta t$ is chosen sufficiently small, there is a variety of 
methods to accurately evaluate the action of the matrix 
exponential $e^{-i{\cal H}\delta t}$ without diagonalization of ${\cal H}$. 
A particularly simple approach in this context is a fourth-order Runge-Kutta 
(RK4) scheme, where the time evolution is approximated as 
\cite{Steinigeweg2014, Elsayed2013},
\begin{equation}\label{Eq::RK1}
 \ket{\psi_\beta(t+\delta t)} \approx \ket{\psi_\beta(t)} + \sum_{k=1}^4 
\ket{f_k}\ .
\end{equation}
The four auxiliary states $\ket{f_1}$ - $\ket{f_4}$ are constructed 
according to \cite{Steinigeweg2014, Elsayed2013},
\begin{equation}\label{Eq::RK2}
 \ket{f_k} =
\frac{-i {\cal H}\delta t}{k} \ket{f_{k-1}}\ ,\quad\ \ket{f_0} = 
\ket{\psi_\beta(t)}\ ,   
\end{equation}
and the error of the approximation \eqref{Eq::RK1} scales as 
${\cal O}(\delta 
t^5)$. Note that the RK4 scheme in Eqs.\ \eqref{Eq::RK1} 
and \eqref{Eq::RK2} is equivalent to a Taylor expansion of the exponential 
$e^{-i{\cal H}\delta t}$ up to fourth order. Note further that, in complete 
analogy to the propagation in real time, the temperature dependence 
of $\ket{\psi_\beta}$ can be generated by an evolution in small 
imaginary time steps $i\delta\beta$. 

Apart from RK4, other common and more sophisticated methods to propagate pure states without 
diagonalization are, e.g., 
Trotter decompositions \cite{DeVries1993,DeRaedt2006}, Krylov subspace techniques 
\cite{Nauts1983}, as well as 
Chebyshev polynomial 
expansions \cite{TalEzer1984, Kosloff1994, Dobrovitski2003, Weisse2006, Kosloff2019}. 
A unifying property of all these 
methods and RK4 is the 
necessity to 
calculate matrix-vector products, i.e., to evaluate the action of the 
Hamiltonian ${\cal H}$ onto pure states.
Importantly, such matrix-vector multiplications can be carried out relatively 
memory efficiently thanks to the sparse matrix structure of ${\cal H}$ in 
models with short-range interactions such as nearest-neighbor couplings. 
As a consequence, it is possible to numerically treat comparatively large system 
sizes, i.e., with huge Hilbert-space dimensions far beyond the range of exact 
diagonalization. 

\subsection{Calculating the (local) density of states}\label{Sec::DOS}
   
As a first useful application, let us describe how pure states, in 
combination with a forward propagation in real time, can be used to evaluate 
the (local) density of states \cite{Hams2000}. To begin with, we note that the 
density of 
states of some Hamiltonian ${\cal H}$ with eigenvalues $E_n$ can be written as
\begin{align}\label{DOS_Eq}
\Omega(E) &= \sum_n \delta(E-E_n)\  \\ &= \frac{1}{2\pi}\int \limits_{ 
-\infty}^\infty e^{itE}\ \text{Tr} [ e^{-i\mathcal{H}t} ] \, \text{d}t\ 
\label{Eq::DOSEq222},
\end{align}
where we have used the definition of the $\delta$ function. 
In the spirit of Eq.\ \eqref{Eq::DQTApprox}, we can approximate the trace in Eq.
\eqref{Eq::DOSEq222} by a scalar product with a randomly drawn pure state 
$\ket{\psi}$, 
\begin{align}\label{Typ_Eq}
\text{Tr} [ e^{-i \mathcal{H}t} ] \propto \bra{\psi} e^{-i\mathcal{H}t}
\ket{\psi} = \braket{\psi|\psi(t)}\ ,
\end{align}
such that Eq.\ \eqref{Eq::DOSEq222} can be approximated as
\begin{equation}\label{DOSEq2}
\Omega(E) \propto \int_{-t_\text{max}}^{+t_\text{max}} e^{itE} 
\braket{\psi|\psi(t)}
\text{d}t\ ,
\end{equation}
where $\braket{\psi(0)|\psi(-t)} = 
\braket{\psi(0)|\psi(t)}^\ast$, and $t_\text{max}$ is the maximum time to 
which $\ket{\psi(t)}$ is evolved. Due to this cutoff time, the resulting energy 
resolution of $\Omega(E)$ is given by $\Delta E = \pi/ 
t_\text{max}$. Thus, the density of states of some Hamiltonian ${\cal H}$ can 
be obtained from the Fourier transform of the survival probability 
$\braket{\psi|\psi(t)}$ of a random pure state \cite{Hams2000}. 

In fact, the relation \eqref{DOSEq2} turns out to be useful 
for any arbitrary pure state $\ket{\tilde{\psi}}$ (which is not 
necessarily drawn at random).
The local density of states $P(E)$ of $\ket{\tilde{\psi}}$, i.e., the 
spectral distribution of $\ket{\tilde{\psi}}$, is then defined as
\begin{align}
P(E) &= \sum_n |\braket{n|\tilde{\psi}}|^2 \, \delta(E-E_n)\ , 
\end{align}
where $\ket{n}$ are the eigenvectors of $\mathcal{H}$ with corresponding 
eigenvalues $E_n$. Analogous to Eq.\ \eqref{DOSEq2}, $P(E)$ can be written as 
the Fourier transform of the survival probability of $\ket{\tilde{\psi}}$ 
\cite{Hams2000, Schiulaz2019}, 
\begin{align}
 P(E) \propto \int_{-t_\text{max}}^{+t_\text{max}} e^{itE} 
\braket{\tilde{\psi}|
\tilde{\psi}(t)}  \text{d}t\ \label{PE_EQ}\ .
\end{align}
Relying on the forward propagation of pure states discussed in Sec.\ 
\ref{Sec::PSP}, it is thus possible to access $\Omega(E)$ and $P(E)$. 
Note that Eqs.\ 
\eqref{DOSEq2} and \eqref{PE_EQ} only provide the overall shape (within 
the resolution $\Delta E$) of $\Omega(E)$ 
and $P(E)$, while single eigenstates are difficult to resolve 
\cite{Richter2018a, Kennes2020}.

\begin{figure}[tb]
 \centering
 \includegraphics[width=\columnwidth]{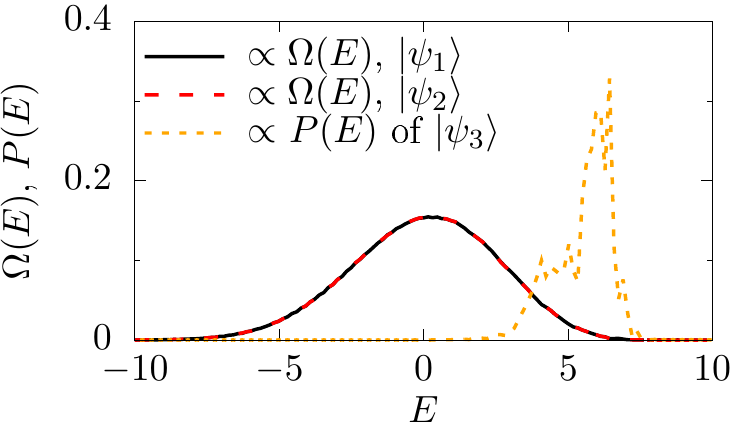}
 \caption{(Color online) Density of states $\Omega(E)$ of a spin-$1/2$ XXZ 
chain with $\Delta = 1.5$ and $L = 24$ sites, obtained from two independently 
drawn random states $\ket{\psi_1}$ and $\ket{\psi_2}$.  
The local density of states $P(E)$ is shown for a nonrandom state 
$\ket{\psi_3}$.
Data is adapted from \cite{Richter2018a}.}
 \label{Fig1}
\end{figure}

As an example, let us consider the 
spin-$1/2$ XXZ chain,
\begin{align}\label{Eq::XXZmodel}
\mathcal{H}=J\sum_{\ell =1 
}^{L}\left(S_{\ell}^{x}S_{\ell+1}^{x}+S_{\ell}^{y}S_{\ell+1}^{y}+\Delta 
S_{\ell}^{z}S_{\ell+1}^{z}\right)\ , 
\end{align}
where $S^{\alpha}_{\ell}$, $\alpha\in\{x,y,z\}$ are the components of
the corresponding spin-$1/2$ operators at the site $\ell$, $L$ is the number of 
lattice sites, 
$J=1$ describes the antiferromagnetic coupling constant, and $\Delta > 0$ is 
the anisotropy in the $z$-direction. In Fig.\ \ref{Fig1}, the 
density of states $\Omega(E)$ is shown for the XXZ chain \eqref{Eq::XXZmodel} 
with $L = 24$ and 
$\Delta = 1.5$, obtained via Eq.\ \eqref{DOSEq2} with two independently drawn 
random states $\ket{\psi_1}$ and $\ket{\psi_2}$. As can be seen in Fig.\ 
\ref{Fig1}, $\Omega(E)$ has a broad 
and Gaussian shape. Moreover, $\Omega(E)$ is essentially the same for the 
two random states, which confirms the accuracy of the typicality approach. In 
addition, we show $P(E)$ for a nonrandom 
state $\ket{\psi_3}$, which is 
sharply peaked at the borders of the spectrum \cite{Richter2018a}. 

\subsection{Time-dependent equilibrium correlation functions}\label{Sec::Trans}

Let us now turn to quantum many-body dynamics within the framework of 
linear response theory (LRT).
Within LRT, central quantities of interest are time-dependent 
correlation functions $C_{AB}(t)$ of two operators
$A$ and $B$
evaluated in equilibrium, 
\begin{align}\label{Eq::autocorrelationfunction}
C_{AB}(t)=
\langle A(t)B\rangle_{\text{eq}}=
\frac{\text{Tr}\left[A(t)Be^{-\beta\mathcal{H}}\right]}{\mathcal{Z}}\ ,
\end{align}
where $\mathcal{Z}$ is again the canonical partition function as defined in 
Eq.\ \eqref{Eq::Equilibrium}, 
and $A(t)$ is the time-evolved operator in the Heisenberg picture.
Analogous to Eq.\ \eqref{Eq::DQTApprox}, $C_{AB}(t)$ can be rewritten according 
to \cite{Iitaka2003, Elsayed2013, Steinigeweg2014},
\begin{align}\label{Eq::DQTCorr}
\langle A(t)B\rangle_{\text{eq}}\approx\frac{\langle 
	\psi_{\beta}(t)|A|\varphi_{\beta}(t)\rangle}{\langle\psi_{\beta}(0) 
	|\psi_{\beta}(0)\rangle}\ , 
\end{align}
where we have introduced two auxiliary pure states,
\begin{align}
\ket{\varphi_{\beta}(t)}&=e^{-i\mathcal{H}t}Be^{-\beta\mathcal{H}/2}\ket{\psi}\ 
, \label{Eq::State11}\\
\ket{\psi_{\beta}(t)}&=e^{-i\mathcal{H}t}e^{-\beta\mathcal{H}/2}\ket{\psi}\ 
\label{Eq::State22},
\end{align}
and $\ket{\psi}$ is a random state drawn from the full Hilbert 
space, cf.\ Eq.\ \eqref{Eq::RandState}. Importantly, in contrast to Eq.\
\eqref{Eq::autocorrelationfunction}, the time (and temperature) argument 
in Eq.\ \eqref{Eq::DQTCorr} is now 
a property of the pure states and not of the operators anymore. According to, 
e.g., Eq.\  
\eqref{Eq::RK1},
$\ket{\varphi_{\beta}(t)}$ and $\ket{\psi_{\beta}(t)}$ can be
evolved in real (and imaginary) time. 

In the context of transport, an interesting quantity is the current 
autocorrelation function $C_{jj}(t)$, which is defined according
to Eq.\ \eqref{Eq::autocorrelationfunction} with 
$A=B=j$, where $j$ is the current operator. Note that the Fourier transform of 
$C_{jj}(t)$ is
related to the conductivity via the
Kubo formula \cite{Kubo1991, Heidrich-Meisner2007}.
\begin{figure}[tb]
 \centering
 \includegraphics[width=\columnwidth]{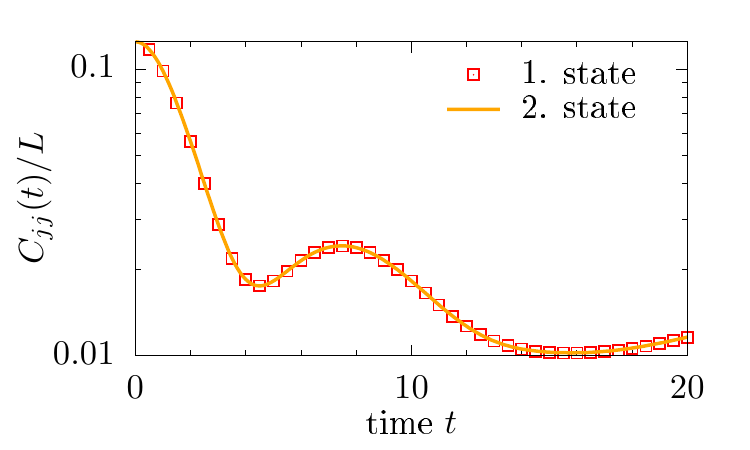}
 \caption{(Color online) Current autocorrelation function $C_{jj}(t)$ at $\beta 
= 0$ for the spin-$1/2$ XXZ chain with $\Delta = 1$, obtained by DQT for $L = 
33$ sites. The calculation is done for two independently drawn states (from 
the symmetry subsector with momentum $k = 0$). Data is adapted from 
Ref.~\cite{Steinigeweg2015}.}
 \label{Fig2}
\end{figure}

For concreteness, let us (again) 
consider the XXZ chain \eqref{Eq::XXZmodel}. In this case, the spin current operator 
$j$ takes on the 
form \cite{Heidrich-Meisner2007},
\begin{align}
j=J\sum_{\ell =1 
}^{L}\left(S_{\ell}^{x}S_{\ell+1}^{y}-S_{\ell}^{y}S_{\ell+1}^{x}\right)\ .
\end{align}
In Refs.\ \cite{Steinigeweg2014, Steinigeweg2015}, 
$C_{jj}(t)$ was studied by means of DQT for the 
XXZ chain with particular focus on
infinite temperature $\beta = 0$.
This infinite-temperature current autocorrelation function is exemplarily shown 
in Fig.\ \ref{Fig2} for $\Delta = 1$ and $L = 33$. To demonstrate the smallness 
of the statistical error of DQT, we show results obtained from two independently 
drawn random states. As can be seen in Fig.\ \ref{Fig2}, both curves coincide 
very well with each other for this choice of $\beta$ and $L$, even in the 
semi-logarithmic plot used.   
(For further numerical data of $C_{jj}(t)$ see also 
Fig.\ \ref{Fig8} below.)

In addition to the XXZ chain, DQT has been used to study $C_{jj}(t)$ for a 
variety of other low-dimensional systems, such as spin chains with 
next-nearest neighbor interactions \cite{Richter2018a} and with spin 
quantum number $S > 1/2$ \cite{Richter2019g}, spin ladders 
\cite{Steinigeweg2014b, Steinigeweg2016, Richter2019f} (also for energy 
currents), as well as Fermi-Hubbard 
chains \cite{Jin2015}. The possibility to 
calculate $C_{jj}(t)$ by means of DQT for large systems and long time scales 
has proven to be very useful to extract transport coefficients, including 
(the finite-size scaling of) dc conductivities, diffusion constants, and 
Drude weights, for integrable and nonintegrable models \cite{Steinigeweg2014, 
Steinigeweg2014b, Jin2015, Steinigeweg2015, Steinigeweg2016a, Steinigeweg2016, 
Richter2018a, Richter2019f, Richter2019g}.

Another interesting quantity in the context of transport are 
the spatio-temporal correlation functions $C_{\ell ,\ell^{\prime}}(t)$ of, e.g.,
spin, which are defined
according to Eq.\ \eqref{Eq::autocorrelationfunction} with
$A=S_{\ell}^{z}$ and $B=S_{\ell^{\prime}}^{z}$,
\begin{align}
C_{\ell ,\ell^\prime}(t)=\langle S_{\ell}^{z}(t)S_{\ell 
	^\prime}^{z}\rangle{_\text{eq}}\ \ . 
\end{align}
While a calculation of $C_{\ell ,\ell^{\prime}}(t)$ can be 
done according to Eq.\ \eqref{Eq::DQTCorr}, a simplification is possible at 
infinite temperature $\beta=0$.
Namely, at $\beta=0$, one can introduce the pure state 
\cite{Richter2019f}
\begin{align}\label{Eq::StateCC}
\ket{\psi^\prime(0)}=\frac{\sqrt{S^{z}_{\ell^{\prime}}+c}\ket{
		\psi}}{ \sqrt { \langle \psi | \psi\rangle}}\ ,
\end{align}
where $\ket{\psi}$ is again drawn randomly according to Eq.\ \eqref{Eq::RandState}, 
and the constant $c$ is chosen such that $S_\ell^z + c$ has 
non-negative eigenvalues. Using Eq.\ \eqref{Eq::StateCC}, one finds
\begin{align}
C_{\ell,\ell^\prime}(t)\ \approx\ \langle 
\psi^{\prime}(t)|S_{\ell}^{z}|\psi^{\prime}(t)\rangle\ .
\end{align}
Thus, it is possible to calculate $C_{\ell ,\ell^{\prime}}(t)$ just 
from one auxiliary state \cite{Richter2019g}, in contrast to 
the current autocorrelations $C_{jj}(t)$, where two states have to be evolved 
in time, cf.\ Eqs.\ \eqref{Eq::State11} and \eqref{Eq::State22}.
\begin{figure}[tb]
\centering
\includegraphics[width=\columnwidth]{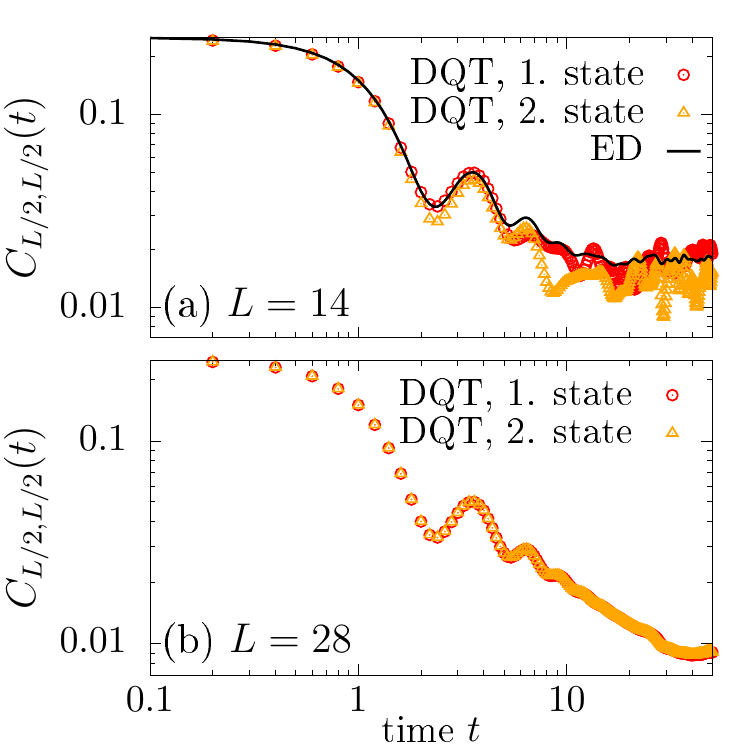}
\caption{(Color online) Equal-site spin-spin correlation function 
$C_{L/2,L/2}(t)$ 
for spin-$1/2$ XXZ chains ($\Delta = 1$) with (a) $L = 14$ 
sites and 
(b) 
$L = 28$ 
sites. For $L = 14$, exact diagonalization 
is compared to DQT for two different random pure 
states. While ED is unfeasible for $L = 28$, the statistical 
fluctuations 
of the typicality approximation become negligible for this 
system size. 
Data is adapted from Ref.\ \cite{Balz2019}.}
	\label{Fig3}
\end{figure}

As an example, the equal-site spin-spin correlation function
$C_{L/2,L/2}(t)$ at lattice site $\ell = L/2$ is shown in Fig.\ \ref{Fig3} for 
spin-$1/2$ XXZ chains 
with two different lengths $L = 14$ and $L = 28$ \cite{Balz2019}. (Note that 
due to periodic boundary conditions, the specific lattice site $\ell$ is 
arbitrary.) As a 
demonstration of the 
accuracy of the DQT approach, the calculation is done for two 
independently drawn pure states $\ket{\psi}$. While the DQT data closely 
follows the exact result at $L = 14$, the residual statistical fluctuations 
disappear almost completely for $L = 28$. Note that while we have 
chosen the XXZ chain to demonstrate the accuracy of DQT for $C_{jj}(t)$ [Fig.\ 
\ref{Fig2}] and for $C_{L/2,L/2}(t)$ [Fig.\ \ref{Fig3}], similar curves can 
be obtained for other models and observables as well. For additional comparisons between DQT data and exact ensemble averages, see, e.g.,\ Refs.\ \cite{Steinigeweg2014,Steinigeweg2014a}. 

\begin{figure}[tb]
	\centering
	\includegraphics[width=\columnwidth]{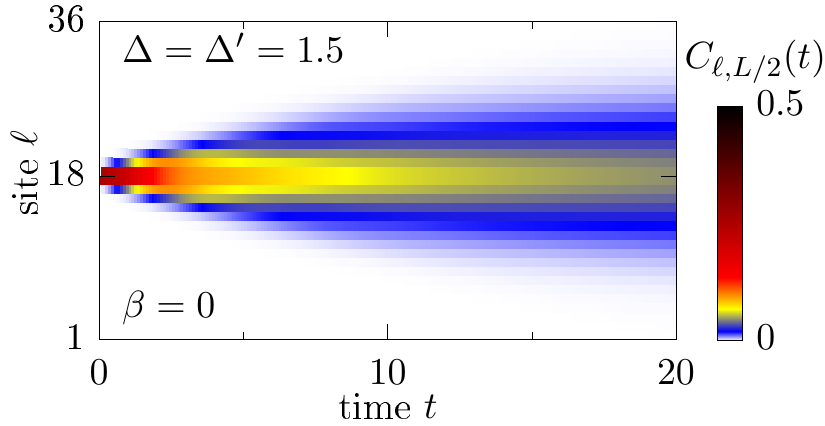}
	\caption{(Color online) Time-space plot of the infinite-temperature 
		spin-spin 
		correlation function 
		$C_{\ell,L/2}(t) = \langle S_\ell^z(t) S_{L/2}^z 
\rangle_\text{eq}$ 
		for a spin-$1/2$ XXZ chain of length $L = 36$, nearest neighbor 
		($\Delta = 1.5$) and next-nearest neighbor ($\Delta' = 1.5$) 
		coupling.  Data is adapted from 
		Ref.\ \cite{Richter2018a}.}
	\label{Fig4}
\end{figure}

As another example, the full time-space profile $C_{\ell,L/2}(t)$ is shown in 
\figref{Fig4} for a 
spin-$1/2$ XXZ chain with next-nearest 
neighbor interactions and $L = 36$ sites \cite{Richter2018a}. While at $\beta=0$
different lattice sites are uncorrelated at $t =0$, correlations 
start to build up for $t > 0$.

\begin{figure}[tb]
\centering
\includegraphics[width=\columnwidth]{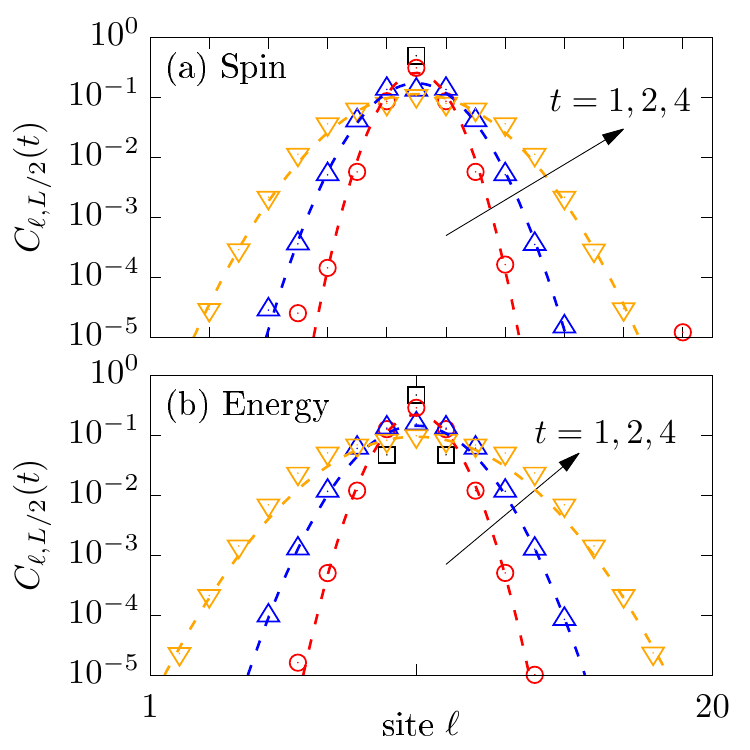}
\caption{(Color online) Spin-spin correlation function 
$C_{\ell,L/2}(t)$ at fixed times, $t = 0$ ($\delta$ peak) and 
$t = 
1,2,4$ 
(arrow),
for a spin-$1/2$ Heisenberg ladder of length $L = 20$ (i.e.\ 
$40$ lattice sites), at high temperatures $\beta = 0$. 
Dashed lines are 
Gaussian fits to the data. Panel (a) shows spin 
densities, while panel (b) shows local energies. 
Data is adapted from Ref.\ \cite{Richter2019f}.}
\label{Fig5}
\end{figure}

A very similar example is shown in Fig.\ \ref{Fig5}, where 
the spatio-temporal correlations for spin and energy 
densities are depicted at fixed times. Yet, the model is a spin-$1/2$ Heisenberg 
ladder,
\begin{equation}
 {\cal H} = J_\parallel \sum_{l=1}^L \sum_{k=1}^2 {\bf S}_{l,k} \cdot {\bf 
S}_{l+1,k}\ + J_\perp \sum_{l = 1}^L {\bf S}_{l,1} \cdot {\bf 
S}_{l,2}\ ,
\end{equation}
where $J_\parallel$ ($J_\perp$) denotes the coupling on the legs (rungs). 
The data in Fig.\ \ref{Fig5} are obtained for $J_\parallel = J_\perp = 1$ and 
$L = 20$, i.e., $40$ lattice sites in total \cite{Richter2019f}. For all times 
shown in Fig.\ 
\ref{Fig5}, one finds that the profiles $C_{\ell,L/2}(t)$ are convincingly 
described by Gaussians, which illustrates once again the 
high accuracy of the DQT approach in the semi-logarithmic plot used. 
Such a Gaussian spreading has been interpreted as a clear signature of 
high-temperature spin and energy diffusion in this and other models 
\cite{Steinigeweg2017, Steinigeweg2017a, Richter2019g, Richter2018a, 
Ljubotina2017}.

In addition, DQT has been used to obtain spatio-temporal 
correlation functions $C_{\ell, \ell^{\prime}}(t)$ in a number of other models. 
Remarkably, clean Gaussian profiles have been found in various parameter 
regimes, even for integrable models such as the spin-$1/2$ XXZ 
chain \cite{Steinigeweg2017a} or the one-dimensional Fermi-Hubbard model  
\cite{Steinigeweg2017}. Other classes of models which have been studied 
in this way include the spin-$1$ XXZ chain \cite{Richter2019g} as well as spin 
models with quenched disorder \cite{Richter2019g, Richter2020}.

\subsection{Applications to far-from-equilibrium dynamics}\label{Sec::FFED}

Nonequilibrium scenarios in isolated quantum systems can be induced via 
explicitly time-dependent Hamiltonians or, 
e.g., by means of quantum quenches \cite{Mitra2018}.
For instance, the system can  
be initially in an eigenstate of some Hamiltonian $\mathcal{H}_1$ while the 
subsequent dynamics are governed by a different Hamiltonian $\mathcal{H}_2$. 

\begin{figure}[tb]
	\centering
		{\includegraphics[width=\columnwidth]{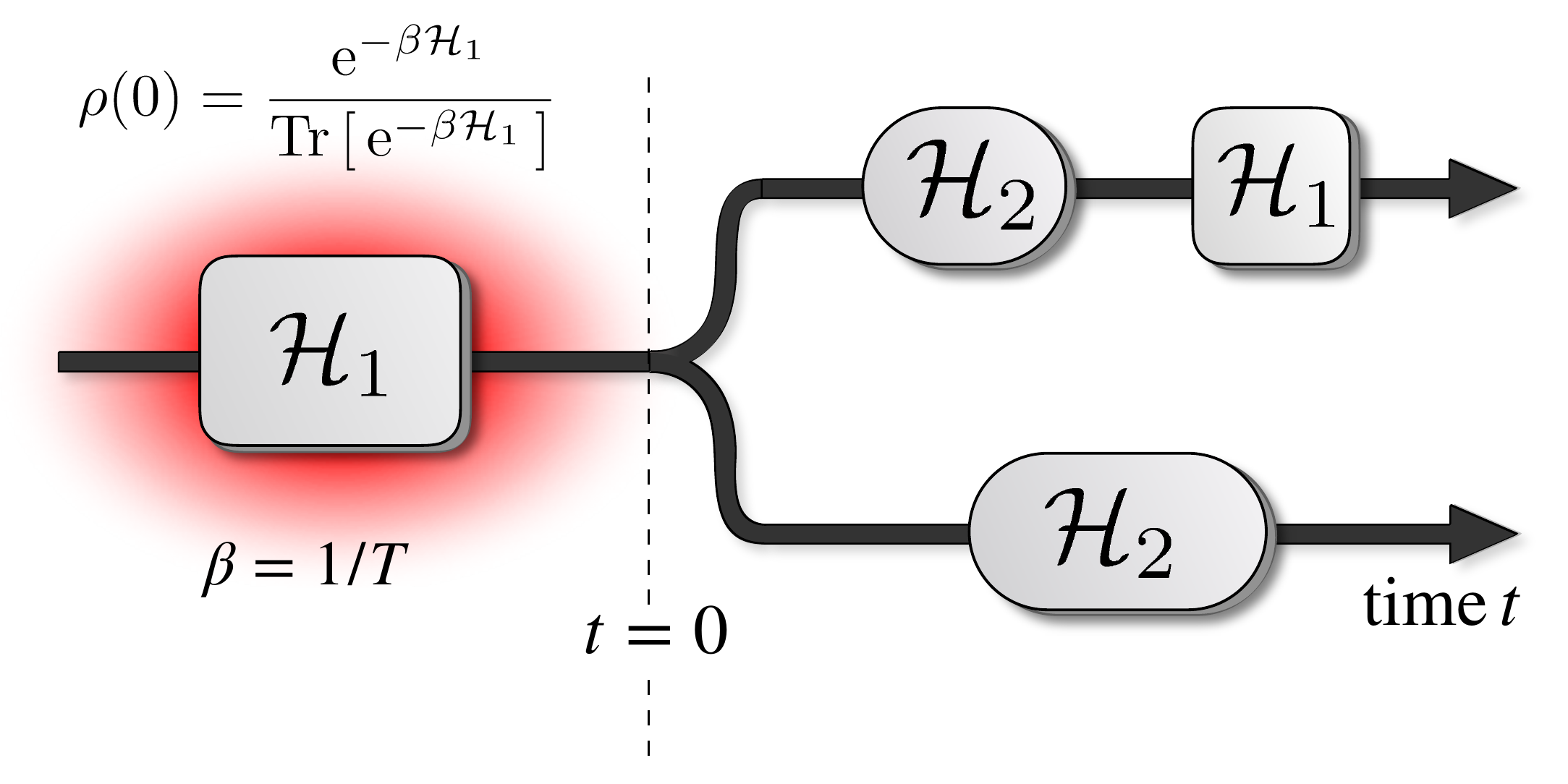}}
	\caption{(Color online) Sketch of the quench protocol. The system starts
	in a Gibbs state with respect to some initial Hamiltonian 
$\mathcal{H}_1$. 
	For times $t>0$, the system evolves unitarily according to some other 
	Hamiltonian $\mathcal{H}_2$ as per $\rho(t)=\mathrm{e}^{-i\mathcal{H}_2t}\,
	\rho(0)\,\mathrm{e}^{i\mathcal{H}_2t}$. This protocol can also be 
	modified by switching back to the original Hamiltonian $\mathcal{H}_1$
	(shown in the upper branch on the right hand side) or by further
	 changes of the Hamiltonian in time.}
    \label{Fig6}
\end{figure}

Here, we discuss an alternative type of 
quench, where the system starts in a Gibbs state with respect to (w.r.t.) some initial 
Hamiltonian $\mathcal{H}_1$ (see Fig.\ \ref{Fig6}), 
\begin{align}\label{Eq::initial-thermal-Gibbs-state}
    \rho(0)=\frac{\emath{-\beta\mathcal{H}_1}}{\mathcal{Z}}\ . 
\end{align}
We then consider a quantum quench, where $\mathcal{H}_1$ is changed to some 
other Hamiltonian $\mathcal{H}_2$. The system then is in a nonequilibrium state 
and evolves unitarily according to the new Hamiltonian,
\begin{align}    
\rho(t)=\timevolve{\rho(0)}{i\mathcal{H}
_2t } \ .
\end{align}
The post-quench Hamiltonian can, for instance, be created by adding or removing 
a static (weak or strong) force of strength $\epsilon$ to the initial 
Hamiltonian, i.e., 
$\mathcal{H}_2=\mathcal{H}_1\pm\epsilon A$, where the 
operator $A$ is conjugated to the force 
\cite{Bartsch2017,Richter2018, Richter2019a, Richter2019c}. 
The resulting expectation value dynamics of, e.g., the operator 
$A$ is given by
\begin{align}\label{Eq::noneq-expectationvalue}
    \braket{A(t)}=\Tr{\rho(t)A}\ ,
\end{align}
and its evaluation in principle requires complete diagonalization of 
both ${\cal H}_1$ and ${\cal H}_2$.
As before, this diagonalization can be circumvented by preparing a 
typical pure state \cite{Bartsch2017,Richter2018, Richter2019a, 
Richter2019c,Endo2018},
\begin{align}\label{Eq::noneq-pure-initial-state}
\ket{\Psi(0)}=\frac{\emath{-\beta\mathcal{H}_1/2}\ket{\psi}}{
\sqrt { \bra {\psi}
    \emath{-\beta\mathcal{H}_1}\ket{\psi}}}\ , 
\end{align}
which mimics the density matrix 
\eqref{Eq::initial-thermal-Gibbs-state}, and 
the reference state $\ket{\psi}$ is again randomly drawn from the 
full Hilbert space, cf.\ Eq.\ \eqref{Eq::RandState}. Both the 
imaginary-time evolution w.r.t.\ $\mathcal{H}_1$ and the real-time 
evolution w.r.t.\ $\mathcal{H}_2$ can be done following 
Sec.\ \ref{Sec::PSP}. In this way, one gets 
\begin{equation}
\braket{A(t)} \approx \bra{\Psi(t)}{A} 
\ket{\Psi(t)}\ . 
\end{equation}
It is worth pointing out that the (simple) quench protocol above can be 
modified by additional changes of the Hamiltonian in time. A static force 
switched on at time $t=0$ can, for instance, be removed again at some later 
time $t>0$, see also Fig.\ \ref{Fig6}. Even for such protocols, the additional 
efforts of the DQT approach are minor 
compared to ED, where the diagonalization of multiple Hamiltonians has to be 
carried out.

\begin{figure}[tb]
    \centering
    \includegraphics[width=\columnwidth]{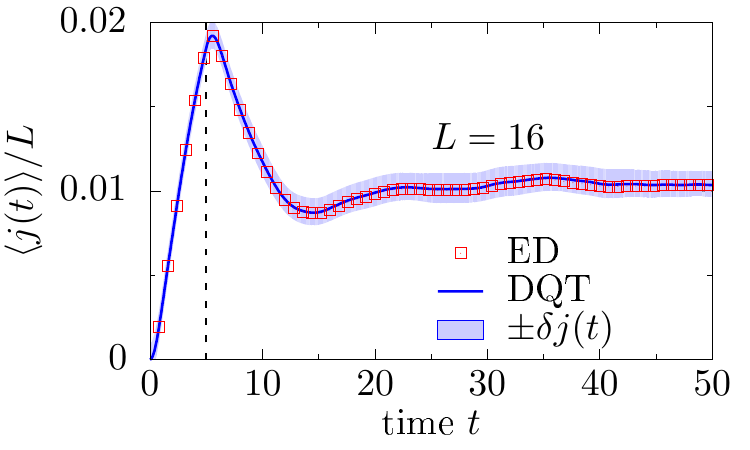}
    \caption{(Color online) Out-of-equilibrium dynamics of the spin current $j$ 
in the spin-$1/2$ XXZ chain with $\Delta = 0.5$ and $L = 16$, starting from a 
thermal state with $\beta = 1$. For times $0< t < 5$, an external force acts on 
the system, which gives rise to an additional term $\propto j$ within 
the Hamiltonian. Results from the typicality approach are compared to exact 
diagonalization. DQT data are averaged over $N = 100$ random initial states and 
the shaded area indicates 
the standard deviation. Data is adapted from 
Ref.\ \cite{Richter2019d}. 
   }
    \label{Fig7}
\end{figure}

In Fig.\ \ref{Fig7}, the nonequilibrium dynamics $\langle 
j(t) \rangle$ of the spin current is 
exemplarily depicted for a XXZ chain which is initially prepared in a thermal 
state at the finite temperature $\beta = 1$ (see caption of Fig.\ \ref{Fig7} and 
Ref.\ \cite{Richter2019d} for a more detailed description of the protocol). 
Here, the accuracy of the DQT approach is demonstrated by comparing to data 
obtained by exact diagonalization.

\subsection{DQT and its extensions}\label{Sec::Ext}

In addition to the direct applications discussed above, 
DQT also is a useful tool to ``boost'' other (numerical or analytical) 
techniques, which can profit from accurate data for large 
system sizes. Two examples of such techniques, which have 
recently been combined with DQT, are numerical linked-cluster expansions 
(NLCE) and projection operator techniques.

\subsubsection{NLCE}
The key feature used in NLCE is the fact that the per-site value of an
extensive quantity on an infinite lattice can be expanded in terms of its 
respective weights on all linked (sub-)clusters that can be embedded in the 
lattice. While NLCE is described in detail and generality in 
\cite{Rigol2006b,Tang2013}, this section focuses on practical 
aspects of NLCE, particularly on its combination with DQT to calculate, e.g., 
current-current correlation functions of one-dimensional systems. 
The starting point of a corresponding NLCE is the expression
\begin{align}\label{Eq::NLCE}
    \frac{\braket{ j(t)j}\subeq}{L}=\sum_c\mathcal{L}_cW_c(t)\ ,
\end{align}
where $W_c$ is the weight of a cluster $c$ with multiplicity $\mathcal{L}_c$. 
To avoid redundant computations, the multiplicity factor (divided by the total 
number of lattice sites) accounts for all clusters,  which are symmetrically or 
topologically related to one representative cluster and therefore yield the same 
weight. The weight of each cluster is evaluated by the inclusion-exclusion 
principle 
\begin{align}
    W_c(t)=\braket{ j(t)j }\subeq^{(c)}-\sum_{s\subset c}W_s(t)\ ,
\end{align}
where the weights of all embedded clusters $s$ are subtracted 
from $\braket{ j(t)j 
}\subeq^{(c)}$ evaluated on the cluster $c$. 

Since the maximum treatable 
cluster size is naturally limited by the available computational resources, 
the sum in Eq. \eqref{Eq::NLCE} has to be truncated to a 
maximum size $c_{\mathrm{max}}$. In one dimension, this truncated sum 
reduces to the difference of the autocorrelation functions of the two largest 
open-boundary chains with length $c_{\mathrm{max}}$ and $c_{\mathrm{max}}-1$, 
i.e.,
\begin{align}\label{Eq::NLCE-recipe}
    \sum_{c=2}^{c_{\mathrm{max}}}W_c(t)=
    \braket{ j(t)j}\subeq^{(c_{\mathrm{max}})}-\braket{ 
j(t)j}\subeq^{(c_{\mathrm{max}}-1)}\ .
\end{align}
As demonstrated in Ref.\ \cite{Richter2019b}, this rather 
simple formula can have a better convergence towards the
thermodynamic limit than a standard finite-size scaling at 
effectively equal computational cost. 

\begin{figure}[tb]
    \centering
    \includegraphics[width=\columnwidth]{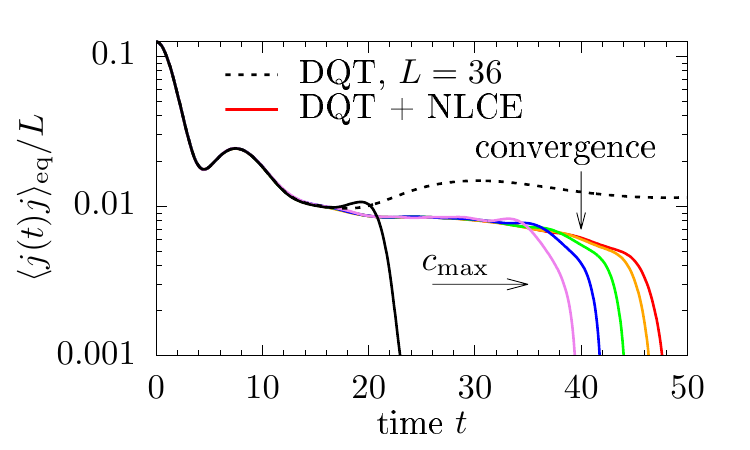}
    \caption{(Color online) Current-current correlation function 
   $\langle j(t) j \rangle_\text{eq}/L$ in the XXZ chain ($\Delta = 1$) at 
$\beta=0$. 
Dashed line indicates data obtained by DQT for $L = 36$ and periodic boundary 
   conditions. Solid lines are obtained by the combination of DQT and NLCE for 
   expansion orders $c_\text{max} = 18, 32, 34, 36, 38, 39$ (arrow).
   Data is adapted from Refs.\ \cite{Richter2019b, Richter2019e}. 
   }
    \label{Fig8}
\end{figure}

As shown in 
\figref{Fig8}, the current autocorrelation function for the spin-1/2 
Heisenberg chain directly obtained by DQT for a large system 
with $L=36$ still exhibits notable finite-size effects 
for times $t > 20$, whereas corresponding DQT+NLCE data is already 
converged for these times. 
Due to the truncation to a maximum cluster size 
$c_\text{max}$, however, the expansion eventually breaks down and only yields 
reliable results up to a maximum time, which increases with 
$c_\text{max}$ \cite{White2017, Mallayya2018, Richter2019b}. 
For the specific example in \figref{Fig8}, this maximum time is 
$t_\text{max} \sim 40$ for the maximum cluster size $c_\text{max}=39$ 
calculated.

When studying thermodynamic quantities, for which the NLCE was originally 
introduced, using larger cluster sizes similarly improves the 
convergence of the expansion down to lower temperatures 
\cite{Tang2013, Bhattaram2019}. 
Either way, it is thus desirable to access cluster sizes as large as possible 
and DQT can be used to
evaluate the contributions of clusters  beyond the range of ED.
Since the difference in Eq. \eqref{Eq::NLCE-recipe} could be 
sensitive to small statistical errors, it might 
be recommended to average the DQT results over multiple random pure states, 
in particular in higher dimensions, where the NLCE expression is not just a 
single difference.

\subsubsection{Projection operator techniques}

The DQT approach can also be used in the context of projection operator 
techniques, e.g., the 
so-called time-convolutionless (TCL) projection operator method. 
These techniques can be applied to situations where a closed quantum system with 
Hamiltonian $\mathcal{H}_0$ is perturbed by an operator
$\mathcal{V}$ with strength $\lambda$, such that the total 
Hamiltonian takes on the form
\begin{align}
    \mathcal{H}=\mathcal{H}_0+\lambda\mathcal{V}\ .
\end{align}
In this setting, one then chooses a suitable projection on the relevant 
degrees of freedom to obtain a systematic perturbation expansion for the
reduced dynamics.
Again, we refer to 
\cite{Chaturvedi1979,Breuer2007,Steinigeweg2011b,Richter2019e} for a 
detailed description of the TCL method and do not discuss it here in full 
generality.
\begin{figure}[tb]
    \centering
    \includegraphics[width=\columnwidth]{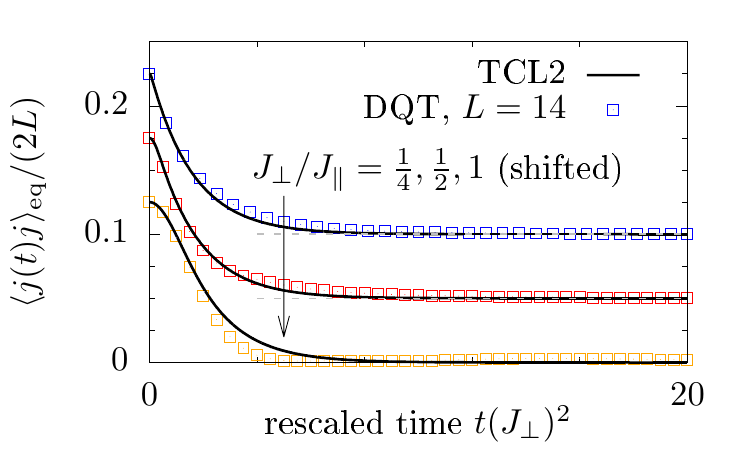}
    \caption{(Color online) Current-current correlation function for spin-$1/2$ 
XX ladders ($J_{\parallel}=1$) with different interchain couplings $J_\perp$ 
(shifted for better 
visibility). Symbols denote exact data obtained by DQT for length $L = 
   14$, i.e., $28$ spins in total. The solid lines indicate the prediction from 
the (second order) TCL projection operator method, cf.\ Eq.\ 
   \eqref{Eq::TCL-Damping}. Data is adapted from Ref.\ \cite{Richter2019e}.
   }
    \label{Fig9}
\end{figure}

Choosing a simple projection onto $A$ only and considering the specific initial
conditions $\rho(0)\sim A$ yields in second order of the perturbation 
\cite{Steinigeweg2011b, Richter2019e}
\begin{align}\label{Eq::TCL-Damping}
\braket{A(t)}_{\mathcal{H}}=\braket{A(t)}_{\mathcal{H}_0}
\exp\left[-\lambda^2 \int_0^t\mathrm{d}t'\gamma_2(t')\right]\ ,
\end{align}
where the second-order damping rate $\gamma_2(t)$ is given by
\begin{align}\label{Eq::2nd-order-kernel}
 \gamma_2(t)=-\int_0^t\ \frac{\mathrm{Tr}\lbrace \commutator{A}{
\mathcal{V}\intpic(t')}\commutator{A}{\mathcal{V}}\rbrace}{\braket{
A^2}} \mathrm{d}t'
\end{align}
and the index $\mathrm{I}$ indicates operators in the interaction picture. 

The calculation of Eq. \eqref{Eq::2nd-order-kernel} can be conveniently done using 
typical states and becomes especially simple in the case where the observable of 
interest is conserved under the unperturbed Hamiltonian, i.e., 
$\commutator{A}{\mathcal{H}_0}=0$.
By defining $\mathcal{K}=\commutator{A}{\mathcal{V}}$ and 
$\mathcal{K}\intpic(t)=\emath{i\mathcal{H}
_0t}\,\mathcal{K}\,\emath{-i\mathcal{H}_0t}$,
the numerator of Eq. \eqref{Eq::2nd-order-kernel} can be calculated as,  
\begin{align}    
\text{Tr}[\mathcal{K}\intpic(t)\mathcal{K}]\propto\braket{\psi(t)\vert\,
\mathcal{K}\,\vert\varphi(t)}\ , 
\end{align}
with the auxiliary states
\begin{align}
   & \ket{\psi(t)}=\emath{-i\mathcal{H}_0t}\ket{\psi}\  \\
   & \ket{\varphi(t)}=\emath{-i\mathcal{H}_0t}\mathcal{K}\ket{\psi} \ .
\end{align} 
In \cite{Richter2019e}, the quality of the second-order 
prediction \eqref{Eq::TCL-Damping} was numerically studied
for the example of the current autocorrelation functions 
$\langle j(t) j\rangle_\text{eq}$ in spin-1/2 ladder systems, 
where the interactions on the rungs of the ladder are treated as a perturbation 
to the 
otherwise uncoupled legs. As depicted in \figref{Fig9}, the 
second-order prediction agrees convincingly with exact data obtained by DQT 
for different strengths of the  
perturbation.

\section{Conclusion}\label{Sec::Sum}

To summarize, we have discussed several applications of dynamical quantum 
typicality and its usefulness as a numerical approach to the real-time dynamics 
of 
quantum many-body systems. The main idea 
of this typicality approach is to approximate ensemble expectation 
values via single pure 
states which are randomly drawn from a high-dimensional Hilbert space. 
In particular, time (temperature) dependencies of expectation values can be 
obtained by iteratively solving the Schr\"odinger equation in 
real (imaginary) time, e.g., 
by means of Runge-Kutta schemes or more sophisticated methods.  

First, we have described that DQT can be used to study the (local) density of 
states as well as equilibrium correlation functions for long time scales and 
comparatively large system sizes beyond the range of standard exact 
diagonalization. Especially in the context of transport, the calculation of 
current autocorrelations and density-density correlations by means of DQT 
is possible. 
Furthermore, we have outlined that DQT is suitable to investigate also
the far-from-equilibrium dynamics resulting from certain quench protocols. 
For instance, an initial Gibbs state is properly imitated 
by a typical pure state and nonequilibrium conditions are induced by removing 
or adding an external force. Eventually, we have discussed that 
DQT can additionally be combined with other approaches. As one example, we have 
shown that the convergence of numerical linked-cluster expansions can be improved 
by evaluating the contributions of larger clusters by means of DQT. As another 
example, we have discussed that DQT allows to compute memory kernels
which arise in projection operator methods such as the TCL technique.  
 
While this paper has illustrated the usefulness of DQT for selected 
applications in the context of transport and thermalization, we should stress that
there certainly are other applications of DQT which have not been mentioned by us. 
One such application, as done in, e.g., \cite{Luitz2017}, is the spreading of quantum information 
measured by so-called out-of-time-ordered correlators (OTOCs) of the form 
\cite{Maldacena2016},
\begin{equation}\label{Eq::OTOC}
 C(t) = \frac{\text{Tr}[A(t) B A(t) B]}{d}\ , 
\end{equation}
where the operators $A$ and $B$ are, for instance, local magnetization densities 
$S_\ell^z$ at two different lattice sites. Similar to the correlation functions 
discussed in Eq.\ \eqref{Eq::autocorrelationfunction}, the OTOC in Eq.\ 
\eqref{Eq::OTOC} can be 
approximated as the overlap $C(t) \approx \braket{\psi_2(t)|\psi_1(t)}$ of the  
two auxiliary states $\ket{\psi_1(t)} = A(t) B \ket{\psi}$ and 
$\ket{\psi_2(t)} = B A(t) \ket{\psi}$, where $\ket{\psi}$ is again a 
Haar-random state \cite{Luitz2017}.    
 
In conclusion, the concept of dynamical quantum typicality offers a rather simple 
yet remarkably useful approach to study the real-time dynamics of 
quantum many-body systems. It is our hope that the examples discussed in this 
paper motivate its application in other areas as well.  

\subsection*{Acknowledgments}

This work has been funded by the 
Deutsche Forschungsgemeinschaft (DFG) - Grants No.\ 397067869 (STE 2243/3-1), 
No.\ 355031190 - within the DFG Research Unit FOR 2692. 


\bibliographystyle{apsrev4-1}
\bibliography{Review}

\begin{thebibliography}{97}%
\makeatletter
\providecommand \@ifxundefined [1]{%
 \@ifx{#1\undefined}
}%
\providecommand \@ifnum [1]{%
 \ifnum #1\expandafter \@firstoftwo
 \else \expandafter \@secondoftwo
 \fi
}%
\providecommand \@ifx [1]{%
 \ifx #1\expandafter \@firstoftwo
 \else \expandafter \@secondoftwo
 \fi
}%
\providecommand \natexlab [1]{#1}%
\providecommand \enquote  [1]{``#1''}%
\providecommand \bibnamefont  [1]{#1}%
\providecommand \bibfnamefont [1]{#1}%
\providecommand \citenamefont [1]{#1}%
\providecommand \href@noop [0]{\@secondoftwo}%
\providecommand \href [0]{\begingroup \@sanitize@url \@href}%
\providecommand \@href[1]{\@@startlink{#1}\@@href}%
\providecommand \@@href[1]{\endgroup#1\@@endlink}%
\providecommand \@sanitize@url [0]{\catcode `\\12\catcode `\$12\catcode
  `\&12\catcode `\#12\catcode `\^12\catcode `\_12\catcode `\%12\relax}%
\providecommand \@@startlink[1]{}%
\providecommand \@@endlink[0]{}%
\providecommand \url  [0]{\begingroup\@sanitize@url \@url }%
\providecommand \@url [1]{\endgroup\@href {#1}{\urlprefix }}%
\providecommand \urlprefix  [0]{URL }%
\providecommand \Eprint [0]{\href }%
\providecommand \doibase [0]{http://dx.doi.org/}%
\providecommand \selectlanguage [0]{\@gobble}%
\providecommand \bibinfo  [0]{\@secondoftwo}%
\providecommand \bibfield  [0]{\@secondoftwo}%
\providecommand \translation [1]{[#1]}%
\providecommand \BibitemOpen [0]{}%
\providecommand \bibitemStop [0]{}%
\providecommand \bibitemNoStop [0]{.\EOS\space}%
\providecommand \EOS [0]{\spacefactor3000\relax}%
\providecommand \BibitemShut  [1]{\csname bibitem#1\endcsname}%
\let\auto@bib@innerbib\@empty
\bibitem [{\citenamefont {Bloch}\ \emph {et~al.}(2012)\citenamefont {Bloch},
  \citenamefont {Dalibard},\ and\ \citenamefont
  {Nascimb{\`{e}}ne}}]{Bloch2012}%
  \BibitemOpen
  \bibfield  {author} {\bibinfo {author} {\bibfnamefont {I.}~\bibnamefont
  {Bloch}}, \bibinfo {author} {\bibfnamefont {J.}~\bibnamefont {Dalibard}}, \
  and\ \bibinfo {author} {\bibfnamefont {S.}~\bibnamefont {Nascimb{\`{e}}ne}},\
  }\href {\doibase 10.1038/nphys2259} {\bibfield  {journal} {\bibinfo
  {journal} {Nat. Phys.}\ }\textbf {\bibinfo {volume} {8}},\ \bibinfo {pages}
  {267} (\bibinfo {year} {2012})}\BibitemShut {NoStop}%
\bibitem [{\citenamefont {Blatt}\ and\ \citenamefont {Roos}(2012)}]{Blatt2012}%
  \BibitemOpen
  \bibfield  {author} {\bibinfo {author} {\bibfnamefont {R.}~\bibnamefont
  {Blatt}}\ and\ \bibinfo {author} {\bibfnamefont {C.~F.}\ \bibnamefont
  {Roos}},\ }\href {\doibase 10.1038/nphys2252} {\bibfield  {journal} {\bibinfo
   {journal} {Nat. Phys.}\ }\textbf {\bibinfo {volume} {8}},\ \bibinfo {pages}
  {277} (\bibinfo {year} {2012})}\BibitemShut {NoStop}%
\bibitem [{\citenamefont {Polkovnikov}\ \emph {et~al.}(2011)\citenamefont
  {Polkovnikov}, \citenamefont {Sengupta}, \citenamefont {Silva},\ and\
  \citenamefont {Vengalattore}}]{Polkovnikov2011}%
  \BibitemOpen
  \bibfield  {author} {\bibinfo {author} {\bibfnamefont {A.}~\bibnamefont
  {Polkovnikov}}, \bibinfo {author} {\bibfnamefont {K.}~\bibnamefont
  {Sengupta}}, \bibinfo {author} {\bibfnamefont {A.}~\bibnamefont {Silva}}, \
  and\ \bibinfo {author} {\bibfnamefont {M.}~\bibnamefont {Vengalattore}},\
  }\href {\doibase 10.1103/RevModPhys.83.863} {\bibfield  {journal} {\bibinfo
  {journal} {Rev. Mod. Phys.}\ }\textbf {\bibinfo {volume} {83}},\ \bibinfo
  {pages} {863} (\bibinfo {year} {2011})}\BibitemShut {NoStop}%
\bibitem [{\citenamefont {Eisert}\ \emph {et~al.}(2015)\citenamefont {Eisert},
  \citenamefont {Friesdorf},\ and\ \citenamefont {Gogolin}}]{Eisert2015}%
  \BibitemOpen
  \bibfield  {author} {\bibinfo {author} {\bibfnamefont {J.}~\bibnamefont
  {Eisert}}, \bibinfo {author} {\bibfnamefont {M.}~\bibnamefont {Friesdorf}}, \
  and\ \bibinfo {author} {\bibfnamefont {C.}~\bibnamefont {Gogolin}},\ }\href
  {\doibase 10.1038/nphys3215} {\bibfield  {journal} {\bibinfo  {journal} {Nat.
  Phys.}\ }\textbf {\bibinfo {volume} {11}},\ \bibinfo {pages} {124} (\bibinfo
  {year} {2015})}\BibitemShut {NoStop}%
\bibitem [{\citenamefont {D'Alessio}\ \emph {et~al.}(2016)\citenamefont
  {D'Alessio}, \citenamefont {Kafri}, \citenamefont {Polkovnikov},\ and\
  \citenamefont {Rigol}}]{Dalessio2016}%
  \BibitemOpen
  \bibfield  {author} {\bibinfo {author} {\bibfnamefont {L.}~\bibnamefont
  {D'Alessio}}, \bibinfo {author} {\bibfnamefont {Y.}~\bibnamefont {Kafri}},
  \bibinfo {author} {\bibfnamefont {A.}~\bibnamefont {Polkovnikov}}, \ and\
  \bibinfo {author} {\bibfnamefont {M.}~\bibnamefont {Rigol}},\ }\href
  {\doibase 10.1080/00018732.2016.1198134} {\bibfield  {journal} {\bibinfo
  {journal} {Adv. Phys.}\ }\textbf {\bibinfo {volume} {65}},\ \bibinfo {pages}
  {239} (\bibinfo {year} {2016})}\BibitemShut {NoStop}%
\bibitem [{\citenamefont {Gogolin}\ and\ \citenamefont
  {Eisert}(2016)}]{Gogolin2016}%
  \BibitemOpen
  \bibfield  {author} {\bibinfo {author} {\bibfnamefont {C.}~\bibnamefont
  {Gogolin}}\ and\ \bibinfo {author} {\bibfnamefont {J.}~\bibnamefont
  {Eisert}},\ }\href {\doibase 10.1088/0034-4885/79/5/056001} {\bibfield
  {journal} {\bibinfo  {journal} {Rep. Prog. Phys.}\ }\textbf {\bibinfo
  {volume} {79}},\ \bibinfo {pages} {056001} (\bibinfo {year}
  {2016})}\BibitemShut {NoStop}%
\bibitem [{\citenamefont {Borgonovi}\ \emph {et~al.}(2016)\citenamefont
  {Borgonovi}, \citenamefont {Izrailev}, \citenamefont {Santos},\ and\
  \citenamefont {Zelevinsky}}]{Borgonovi2016}%
  \BibitemOpen
  \bibfield  {author} {\bibinfo {author} {\bibfnamefont {F.}~\bibnamefont
  {Borgonovi}}, \bibinfo {author} {\bibfnamefont {F.}~\bibnamefont {Izrailev}},
  \bibinfo {author} {\bibfnamefont {L.~F.}\ \bibnamefont {Santos}}, \ and\
  \bibinfo {author} {\bibfnamefont {V.}~\bibnamefont {Zelevinsky}},\ }\href
  {\doibase 10.1016/j.physrep.2016.02.005} {\bibfield  {journal} {\bibinfo
  {journal} {Phys. Rep.}\ }\textbf {\bibinfo {volume} {626}},\ \bibinfo {pages}
  {1} (\bibinfo {year} {2016})}\BibitemShut {NoStop}%
\bibitem [{\citenamefont {Buchanan}(2005)}]{Buchanan2005}%
  \BibitemOpen
  \bibfield  {author} {\bibinfo {author} {\bibfnamefont {M.}~\bibnamefont
  {Buchanan}},\ }\href {\doibase 10.1038/nphys157} {\bibfield  {journal}
  {\bibinfo  {journal} {Nat. Phys.}\ }\textbf {\bibinfo {volume} {1}},\
  \bibinfo {pages} {71} (\bibinfo {year} {2005})}\BibitemShut {NoStop}%
\bibitem [{\citenamefont {Khemani}\ \emph {et~al.}(2018)\citenamefont
  {Khemani}, \citenamefont {Vishwanath},\ and\ \citenamefont
  {Huse}}]{Khemani2018}%
  \BibitemOpen
  \bibfield  {author} {\bibinfo {author} {\bibfnamefont {V.}~\bibnamefont
  {Khemani}}, \bibinfo {author} {\bibfnamefont {A.}~\bibnamefont {Vishwanath}},
  \ and\ \bibinfo {author} {\bibfnamefont {D.~A.}\ \bibnamefont {Huse}},\
  }\href {\doibase 10.1103/PhysRevX.8.031057} {\bibfield  {journal} {\bibinfo
  {journal} {Phys. Rev. X}\ }\textbf {\bibinfo {volume} {8}},\ \bibinfo {pages}
  {031057} (\bibinfo {year} {2018})}\BibitemShut {NoStop}%
\bibitem [{\citenamefont {Bertini}\ \emph {et~al.}()\citenamefont {Bertini},
  \citenamefont {Heidrich-Meisner}, \citenamefont {Karrasch}, \citenamefont
  {Prosen}, \citenamefont {Steinigeweg},\ and\ \citenamefont
  {{\v{Z}}nidari{\v{c}}}}]{Bertini2020}%
  \BibitemOpen
  \bibfield  {author} {\bibinfo {author} {\bibfnamefont {B.}~\bibnamefont
  {Bertini}}, \bibinfo {author} {\bibfnamefont {F.}~\bibnamefont
  {Heidrich-Meisner}}, \bibinfo {author} {\bibfnamefont {C.}~\bibnamefont
  {Karrasch}}, \bibinfo {author} {\bibfnamefont {T.}~\bibnamefont {Prosen}},
  \bibinfo {author} {\bibfnamefont {R.}~\bibnamefont {Steinigeweg}}, \ and\
  \bibinfo {author} {\bibfnamefont {M.}~\bibnamefont {{\v{Z}}nidari{\v{c}}}},\
  }\href {http://arxiv.org/abs/2003.03334} {\ }\Eprint
  {http://arxiv.org/abs/2003.03334} {arXiv:2003.03334} \BibitemShut {NoStop}%
\bibitem [{\citenamefont {Pr{\"{u}}fer}\ \emph {et~al.}(2018)\citenamefont
  {Pr{\"{u}}fer}, \citenamefont {Kunkel}, \citenamefont {Strobel},
  \citenamefont {Lannig}, \citenamefont {Linnemann}, \citenamefont {Schmied},
  \citenamefont {Berges}, \citenamefont {Gasenzer},\ and\ \citenamefont
  {Oberthaler}}]{Pruefer2018}%
  \BibitemOpen
  \bibfield  {author} {\bibinfo {author} {\bibfnamefont {M.}~\bibnamefont
  {Pr{\"{u}}fer}}, \bibinfo {author} {\bibfnamefont {P.}~\bibnamefont
  {Kunkel}}, \bibinfo {author} {\bibfnamefont {H.}~\bibnamefont {Strobel}},
  \bibinfo {author} {\bibfnamefont {S.}~\bibnamefont {Lannig}}, \bibinfo
  {author} {\bibfnamefont {D.}~\bibnamefont {Linnemann}}, \bibinfo {author}
  {\bibfnamefont {C.-M.}\ \bibnamefont {Schmied}}, \bibinfo {author}
  {\bibfnamefont {J.}~\bibnamefont {Berges}}, \bibinfo {author} {\bibfnamefont
  {T.}~\bibnamefont {Gasenzer}}, \ and\ \bibinfo {author} {\bibfnamefont
  {M.~K.}\ \bibnamefont {Oberthaler}},\ }\href {\doibase
  10.1038/s41586-018-0659-0} {\bibfield  {journal} {\bibinfo  {journal}
  {Nature}\ }\textbf {\bibinfo {volume} {563}},\ \bibinfo {pages} {217}
  (\bibinfo {year} {2018})}\BibitemShut {NoStop}%
\bibitem [{\citenamefont {Erne}\ \emph {et~al.}(2018)\citenamefont {Erne},
  \citenamefont {B{\"{u}}cker}, \citenamefont {Gasenzer}, \citenamefont
  {Berges},\ and\ \citenamefont {Schmiedmayer}}]{Erne2018}%
  \BibitemOpen
  \bibfield  {author} {\bibinfo {author} {\bibfnamefont {S.}~\bibnamefont
  {Erne}}, \bibinfo {author} {\bibfnamefont {R.}~\bibnamefont {B{\"{u}}cker}},
  \bibinfo {author} {\bibfnamefont {T.}~\bibnamefont {Gasenzer}}, \bibinfo
  {author} {\bibfnamefont {J.}~\bibnamefont {Berges}}, \ and\ \bibinfo {author}
  {\bibfnamefont {J.}~\bibnamefont {Schmiedmayer}},\ }\href {\doibase
  10.1038/s41586-018-0667-0} {\bibfield  {journal} {\bibinfo  {journal}
  {Nature}\ }\textbf {\bibinfo {volume} {563}},\ \bibinfo {pages} {225}
  (\bibinfo {year} {2018})}\BibitemShut {NoStop}%
\bibitem [{\citenamefont {Richter}\ and\ \citenamefont
  {Steinigeweg}(2019{\natexlab{a}})}]{Richter2019c}%
  \BibitemOpen
  \bibfield  {author} {\bibinfo {author} {\bibfnamefont {J.}~\bibnamefont
  {Richter}}\ and\ \bibinfo {author} {\bibfnamefont {R.}~\bibnamefont
  {Steinigeweg}},\ }\href {\doibase 10.1103/PhysRevE.99.012114} {\bibfield
  {journal} {\bibinfo  {journal} {Phys. Rev. E}\ }\textbf {\bibinfo {volume}
  {99}},\ \bibinfo {pages} {012114} (\bibinfo {year}
  {2019}{\natexlab{a}})}\BibitemShut {NoStop}%
\bibitem [{\citenamefont {Kubo}\ \emph {et~al.}(1991)\citenamefont {Kubo},
  \citenamefont {Toda},\ and\ \citenamefont {Hashitsume}}]{Kubo1991}%
  \BibitemOpen
  \bibfield  {author} {\bibinfo {author} {\bibfnamefont {R.}~\bibnamefont
  {Kubo}}, \bibinfo {author} {\bibfnamefont {M.}~\bibnamefont {Toda}}, \ and\
  \bibinfo {author} {\bibfnamefont {N.}~\bibnamefont {Hashitsume}},\ }\href
  {\doibase 10.1007/978-3-642-96701-6} {\emph {\bibinfo {title} {{Statistical
  Physics II}}}},\ \bibinfo {edition} {2nd}\ ed.,\ \bibinfo {series} {Springer
  Series in Solid-State Sciences}, Vol.~\bibinfo {volume} {31}\ (\bibinfo
  {publisher} {Springer},\ \bibinfo {address} {Berlin, Heidelberg},\ \bibinfo
  {year} {1991})\ p.\ \bibinfo {pages} {279}\BibitemShut {NoStop}%
\bibitem [{\citenamefont {Schollw{\"{o}}ck}(2011)}]{Schollwoeck2011}%
  \BibitemOpen
  \bibfield  {author} {\bibinfo {author} {\bibfnamefont {U.}~\bibnamefont
  {Schollw{\"{o}}ck}},\ }\href {\doibase 10.1016/j.aop.2010.09.012} {\bibfield
  {journal} {\bibinfo  {journal} {Ann. Phys. (N. Y).}\ }\textbf {\bibinfo
  {volume} {326}},\ \bibinfo {pages} {96} (\bibinfo {year} {2011})}\BibitemShut
  {NoStop}%
\bibitem [{\citenamefont {Paeckel}\ \emph {et~al.}(2019)\citenamefont
  {Paeckel}, \citenamefont {K{\"{o}}hler}, \citenamefont {Swoboda},
  \citenamefont {Manmana}, \citenamefont {Schollw{\"{o}}ck},\ and\
  \citenamefont {Hubig}}]{Paeckel2019}%
  \BibitemOpen
  \bibfield  {author} {\bibinfo {author} {\bibfnamefont {S.}~\bibnamefont
  {Paeckel}}, \bibinfo {author} {\bibfnamefont {T.}~\bibnamefont
  {K{\"{o}}hler}}, \bibinfo {author} {\bibfnamefont {A.}~\bibnamefont
  {Swoboda}}, \bibinfo {author} {\bibfnamefont {S.~R.}\ \bibnamefont
  {Manmana}}, \bibinfo {author} {\bibfnamefont {U.}~\bibnamefont
  {Schollw{\"{o}}ck}}, \ and\ \bibinfo {author} {\bibfnamefont
  {C.}~\bibnamefont {Hubig}},\ }\href {\doibase 10.1016/j.aop.2019.167998}
  {\bibfield  {journal} {\bibinfo  {journal} {Ann. Phys. (N. Y).}\ }\textbf
  {\bibinfo {volume} {411}},\ \bibinfo {pages} {167998} (\bibinfo {year}
  {2019})}\BibitemShut {NoStop}%
\bibitem [{\citenamefont {Bartsch}\ and\ \citenamefont
  {Gemmer}(2009)}]{Bartsch2009}%
  \BibitemOpen
  \bibfield  {author} {\bibinfo {author} {\bibfnamefont {C.}~\bibnamefont
  {Bartsch}}\ and\ \bibinfo {author} {\bibfnamefont {J.}~\bibnamefont
  {Gemmer}},\ }\href {\doibase 10.1103/PhysRevLett.102.110403} {\bibfield
  {journal} {\bibinfo  {journal} {Phys. Rev. Lett.}\ }\textbf {\bibinfo
  {volume} {102}},\ \bibinfo {pages} {110403} (\bibinfo {year}
  {2009})}\BibitemShut {NoStop}%
\bibitem [{\citenamefont {Reimann}(2018)}]{Reimann2018}%
  \BibitemOpen
  \bibfield  {author} {\bibinfo {author} {\bibfnamefont {P.}~\bibnamefont
  {Reimann}},\ }\href {\doibase 10.1103/PhysRevE.97.062129} {\bibfield
  {journal} {\bibinfo  {journal} {Phys. Rev. E}\ }\textbf {\bibinfo {volume}
  {97}},\ \bibinfo {pages} {062129} (\bibinfo {year} {2018})}\BibitemShut
  {NoStop}%
\bibitem [{\citenamefont {Alben}\ \emph {et~al.}(1975)\citenamefont {Alben},
  \citenamefont {Blume}, \citenamefont {Krakauer},\ and\ \citenamefont
  {Schwartz}}]{Alben1975}%
  \BibitemOpen
  \bibfield  {author} {\bibinfo {author} {\bibfnamefont {R.}~\bibnamefont
  {Alben}}, \bibinfo {author} {\bibfnamefont {M.}~\bibnamefont {Blume}},
  \bibinfo {author} {\bibfnamefont {H.}~\bibnamefont {Krakauer}}, \ and\
  \bibinfo {author} {\bibfnamefont {L.}~\bibnamefont {Schwartz}},\ }\href
  {\doibase 10.1103/PhysRevB.12.4090} {\bibfield  {journal} {\bibinfo
  {journal} {Phys. Rev. B}\ }\textbf {\bibinfo {volume} {12}},\ \bibinfo
  {pages} {4090} (\bibinfo {year} {1975})}\BibitemShut {NoStop}%
\bibitem [{\citenamefont {Lloyd}(1988)}]{Lloyd1988}%
  \BibitemOpen
  \bibfield  {author} {\bibinfo {author} {\bibfnamefont {S.}~\bibnamefont
  {Lloyd}},\ }\emph {\bibinfo {title} {{Pure state quantum statistical
  mechanics and black holes}}},\ \href {http://arxiv.org/abs/1307.0378}
  {\bibinfo {type} {Phd thesis}},\ \bibinfo  {school} {The Rockefeller
  University} (\bibinfo {year} {1988}),\ \Eprint
  {http://arxiv.org/abs/1307.0378} {arXiv:1307.0378} \BibitemShut {NoStop}%
\bibitem [{\citenamefont {{De Raedt}}\ and\ \citenamefont
  {de~Vries}(1989)}]{DeRaedt1989}%
  \BibitemOpen
  \bibfield  {author} {\bibinfo {author} {\bibfnamefont {H.}~\bibnamefont {{De
  Raedt}}}\ and\ \bibinfo {author} {\bibfnamefont {P.}~\bibnamefont
  {de~Vries}},\ }\href {\doibase 10.1007/BF01313668} {\bibfield  {journal}
  {\bibinfo  {journal} {Z. Phys. B}\ }\textbf {\bibinfo {volume} {77}},\
  \bibinfo {pages} {243} (\bibinfo {year} {1989})}\BibitemShut {NoStop}%
\bibitem [{\citenamefont {Jakli{\v{c}}}\ and\ \citenamefont
  {Prelov{\v{s}}ek}(1994)}]{Jaklic1994}%
  \BibitemOpen
  \bibfield  {author} {\bibinfo {author} {\bibfnamefont {J.}~\bibnamefont
  {Jakli{\v{c}}}}\ and\ \bibinfo {author} {\bibfnamefont {P.}~\bibnamefont
  {Prelov{\v{s}}ek}},\ }\href {\doibase 10.1103/PhysRevB.49.5065} {\bibfield
  {journal} {\bibinfo  {journal} {Phys. Rev. B}\ }\textbf {\bibinfo {volume}
  {49}},\ \bibinfo {pages} {5065} (\bibinfo {year} {1994})}\BibitemShut
  {NoStop}%
\bibitem [{\citenamefont {Gemmer}\ \emph {et~al.}(2004)\citenamefont {Gemmer},
  \citenamefont {Michel},\ and\ \citenamefont {Mahler}}]{Gemmer2004}%
  \BibitemOpen
  \bibfield  {author} {\bibinfo {author} {\bibfnamefont {J.}~\bibnamefont
  {Gemmer}}, \bibinfo {author} {\bibfnamefont {M.}~\bibnamefont {Michel}}, \
  and\ \bibinfo {author} {\bibfnamefont {G.}~\bibnamefont {Mahler}},\ }\href
  {\doibase 10.1007/b98082} {\emph {\bibinfo {title} {{Quantum
  Thermodynamics}}}},\ \bibinfo {series} {Lecture Notes in Physics}, Vol.\
  \bibinfo {volume} {657}\ (\bibinfo  {publisher} {Springer},\ \bibinfo
  {address} {Berlin, Heidelberg},\ \bibinfo {year} {2004})\BibitemShut
  {NoStop}%
\bibitem [{\citenamefont {Popescu}\ \emph {et~al.}(2006)\citenamefont
  {Popescu}, \citenamefont {Short},\ and\ \citenamefont
  {Winter}}]{Popescu2006}%
  \BibitemOpen
  \bibfield  {author} {\bibinfo {author} {\bibfnamefont {S.}~\bibnamefont
  {Popescu}}, \bibinfo {author} {\bibfnamefont {A.~J.}\ \bibnamefont {Short}},
  \ and\ \bibinfo {author} {\bibfnamefont {A.}~\bibnamefont {Winter}},\ }\href
  {\doibase 10.1038/nphys444} {\bibfield  {journal} {\bibinfo  {journal} {Nat.
  Phys.}\ }\textbf {\bibinfo {volume} {2}},\ \bibinfo {pages} {754} (\bibinfo
  {year} {2006})}\BibitemShut {NoStop}%
\bibitem [{\citenamefont {Goldstein}\ \emph {et~al.}(2006)\citenamefont
  {Goldstein}, \citenamefont {Lebowitz}, \citenamefont {Tumulka},\ and\
  \citenamefont {Zangh{\`{i}}}}]{Goldstein2006}%
  \BibitemOpen
  \bibfield  {author} {\bibinfo {author} {\bibfnamefont {S.}~\bibnamefont
  {Goldstein}}, \bibinfo {author} {\bibfnamefont {J.~L.}\ \bibnamefont
  {Lebowitz}}, \bibinfo {author} {\bibfnamefont {R.}~\bibnamefont {Tumulka}}, \
  and\ \bibinfo {author} {\bibfnamefont {N.}~\bibnamefont {Zangh{\`{i}}}},\
  }\href {\doibase 10.1103/PhysRevLett.96.050403} {\bibfield  {journal}
  {\bibinfo  {journal} {Phys. Rev. Lett.}\ }\textbf {\bibinfo {volume} {96}},\
  \bibinfo {pages} {050403} (\bibinfo {year} {2006})}\BibitemShut {NoStop}%
\bibitem [{\citenamefont {Reimann}(2007)}]{Reimann2007}%
  \BibitemOpen
  \bibfield  {author} {\bibinfo {author} {\bibfnamefont {P.}~\bibnamefont
  {Reimann}},\ }\href {\doibase 10.1103/PhysRevLett.99.160404} {\bibfield
  {journal} {\bibinfo  {journal} {Phys. Rev. Lett.}\ }\textbf {\bibinfo
  {volume} {99}},\ \bibinfo {pages} {160404} (\bibinfo {year}
  {2007})}\BibitemShut {NoStop}%
\bibitem [{\citenamefont {Iitaka}\ and\ \citenamefont
  {Ebisuzaki}(2003)}]{Iitaka2003}%
  \BibitemOpen
  \bibfield  {author} {\bibinfo {author} {\bibfnamefont {T.}~\bibnamefont
  {Iitaka}}\ and\ \bibinfo {author} {\bibfnamefont {T.}~\bibnamefont
  {Ebisuzaki}},\ }\href {\doibase 10.1103/PhysRevLett.90.047203} {\bibfield
  {journal} {\bibinfo  {journal} {Phys. Rev. Lett.}\ }\textbf {\bibinfo
  {volume} {90}},\ \bibinfo {pages} {047203} (\bibinfo {year}
  {2003})}\BibitemShut {NoStop}%
\bibitem [{\citenamefont {Elsayed}\ and\ \citenamefont
  {Fine}(2013)}]{Elsayed2013}%
  \BibitemOpen
  \bibfield  {author} {\bibinfo {author} {\bibfnamefont {T.~A.}\ \bibnamefont
  {Elsayed}}\ and\ \bibinfo {author} {\bibfnamefont {B.~V.}\ \bibnamefont
  {Fine}},\ }\href {\doibase 10.1103/PhysRevLett.110.070404} {\bibfield
  {journal} {\bibinfo  {journal} {Phys. Rev. Lett.}\ }\textbf {\bibinfo
  {volume} {110}},\ \bibinfo {pages} {070404} (\bibinfo {year}
  {2013})}\BibitemShut {NoStop}%
\bibitem [{\citenamefont {Steinigeweg}\ \emph
  {et~al.}(2014{\natexlab{a}})\citenamefont {Steinigeweg}, \citenamefont
  {Gemmer},\ and\ \citenamefont {Brenig}}]{Steinigeweg2014}%
  \BibitemOpen
  \bibfield  {author} {\bibinfo {author} {\bibfnamefont {R.}~\bibnamefont
  {Steinigeweg}}, \bibinfo {author} {\bibfnamefont {J.}~\bibnamefont {Gemmer}},
  \ and\ \bibinfo {author} {\bibfnamefont {W.}~\bibnamefont {Brenig}},\ }\href
  {\doibase 10.1103/PhysRevLett.112.120601} {\bibfield  {journal} {\bibinfo
  {journal} {Phys. Rev. Lett.}\ }\textbf {\bibinfo {volume} {112}},\ \bibinfo
  {pages} {120601} (\bibinfo {year} {2014}{\natexlab{a}})}\BibitemShut
  {NoStop}%
\bibitem [{\citenamefont {Okamoto}\ \emph {et~al.}(2018)\citenamefont
  {Okamoto}, \citenamefont {Alvarez}, \citenamefont {Dagotto},\ and\
  \citenamefont {Tohyama}}]{Okamoto2018}%
  \BibitemOpen
  \bibfield  {author} {\bibinfo {author} {\bibfnamefont {S.}~\bibnamefont
  {Okamoto}}, \bibinfo {author} {\bibfnamefont {G.}~\bibnamefont {Alvarez}},
  \bibinfo {author} {\bibfnamefont {E.}~\bibnamefont {Dagotto}}, \ and\
  \bibinfo {author} {\bibfnamefont {T.}~\bibnamefont {Tohyama}},\ }\href
  {\doibase 10.1103/PhysRevE.97.043308} {\bibfield  {journal} {\bibinfo
  {journal} {Phys. Rev. E}\ }\textbf {\bibinfo {volume} {97}},\ \bibinfo
  {pages} {043308} (\bibinfo {year} {2018})}\BibitemShut {NoStop}%
\bibitem [{\citenamefont {Yamaji}\ \emph {et~al.}()\citenamefont {Yamaji},
  \citenamefont {Suzuki},\ and\ \citenamefont {Kawamura}}]{Yamaji2018}%
  \BibitemOpen
  \bibfield  {author} {\bibinfo {author} {\bibfnamefont {Y.}~\bibnamefont
  {Yamaji}}, \bibinfo {author} {\bibfnamefont {T.}~\bibnamefont {Suzuki}}, \
  and\ \bibinfo {author} {\bibfnamefont {M.}~\bibnamefont {Kawamura}},\ }\href
  {http://arxiv.org/abs/1802.02854} {\ }\Eprint
  {http://arxiv.org/abs/1802.02854} {arXiv:1802.02854} \BibitemShut {NoStop}%
\bibitem [{\citenamefont {Rousochatzakis}\ \emph {et~al.}(2019)\citenamefont
  {Rousochatzakis}, \citenamefont {Kourtis}, \citenamefont {Knolle},
  \citenamefont {Moessner},\ and\ \citenamefont
  {Perkins}}]{Rousochatzakis2019}%
  \BibitemOpen
  \bibfield  {author} {\bibinfo {author} {\bibfnamefont {I.}~\bibnamefont
  {Rousochatzakis}}, \bibinfo {author} {\bibfnamefont {S.}~\bibnamefont
  {Kourtis}}, \bibinfo {author} {\bibfnamefont {J.}~\bibnamefont {Knolle}},
  \bibinfo {author} {\bibfnamefont {R.}~\bibnamefont {Moessner}}, \ and\
  \bibinfo {author} {\bibfnamefont {N.~B.}\ \bibnamefont {Perkins}},\ }\href
  {\doibase 10.1103/PhysRevB.100.045117} {\bibfield  {journal} {\bibinfo
  {journal} {Phys. Rev. B}\ }\textbf {\bibinfo {volume} {100}},\ \bibinfo
  {pages} {045117} (\bibinfo {year} {2019})}\BibitemShut {NoStop}%
\bibitem [{\citenamefont {Hams}\ and\ \citenamefont {{De
  Raedt}}(2000)}]{Hams2000}%
  \BibitemOpen
  \bibfield  {author} {\bibinfo {author} {\bibfnamefont {A.}~\bibnamefont
  {Hams}}\ and\ \bibinfo {author} {\bibfnamefont {H.}~\bibnamefont {{De
  Raedt}}},\ }\href {\doibase 10.1103/PhysRevE.62.4365} {\bibfield  {journal}
  {\bibinfo  {journal} {Phys. Rev. E}\ }\textbf {\bibinfo {volume} {62}},\
  \bibinfo {pages} {4365} (\bibinfo {year} {2000})}\BibitemShut {NoStop}%
\bibitem [{\citenamefont {de~Vries}\ and\ \citenamefont {{De
  Raedt}}(1993)}]{DeVries1993}%
  \BibitemOpen
  \bibfield  {author} {\bibinfo {author} {\bibfnamefont {P.}~\bibnamefont
  {de~Vries}}\ and\ \bibinfo {author} {\bibfnamefont {H.}~\bibnamefont {{De
  Raedt}}},\ }\href {\doibase 10.1103/PhysRevB.47.7929} {\bibfield  {journal}
  {\bibinfo  {journal} {Phys. Rev. B}\ }\textbf {\bibinfo {volume} {47}},\
  \bibinfo {pages} {7929} (\bibinfo {year} {1993})}\BibitemShut {NoStop}%
\bibitem [{\citenamefont {Sugiura}\ and\ \citenamefont
  {Shimizu}(2012)}]{Sugiura2012}%
  \BibitemOpen
  \bibfield  {author} {\bibinfo {author} {\bibfnamefont {S.}~\bibnamefont
  {Sugiura}}\ and\ \bibinfo {author} {\bibfnamefont {A.}~\bibnamefont
  {Shimizu}},\ }\href {\doibase 10.1103/PhysRevLett.108.240401} {\bibfield
  {journal} {\bibinfo  {journal} {Phys. Rev. Lett.}\ }\textbf {\bibinfo
  {volume} {108}},\ \bibinfo {pages} {240401} (\bibinfo {year}
  {2012})}\BibitemShut {NoStop}%
\bibitem [{\citenamefont {Sugiura}\ and\ \citenamefont
  {Shimizu}(2013)}]{Sugiura2013}%
  \BibitemOpen
  \bibfield  {author} {\bibinfo {author} {\bibfnamefont {S.}~\bibnamefont
  {Sugiura}}\ and\ \bibinfo {author} {\bibfnamefont {A.}~\bibnamefont
  {Shimizu}},\ }\href {\doibase 10.1103/PhysRevLett.111.010401} {\bibfield
  {journal} {\bibinfo  {journal} {Phys. Rev. Lett.}\ }\textbf {\bibinfo
  {volume} {111}},\ \bibinfo {pages} {010401} (\bibinfo {year}
  {2013})}\BibitemShut {NoStop}%
\bibitem [{\citenamefont {Wietek}\ \emph {et~al.}(2019)\citenamefont {Wietek},
  \citenamefont {Corboz}, \citenamefont {Wessel}, \citenamefont {Normand},
  \citenamefont {Mila},\ and\ \citenamefont {Honecker}}]{Wietek2019}%
  \BibitemOpen
  \bibfield  {author} {\bibinfo {author} {\bibfnamefont {A.}~\bibnamefont
  {Wietek}}, \bibinfo {author} {\bibfnamefont {P.}~\bibnamefont {Corboz}},
  \bibinfo {author} {\bibfnamefont {S.}~\bibnamefont {Wessel}}, \bibinfo
  {author} {\bibfnamefont {B.}~\bibnamefont {Normand}}, \bibinfo {author}
  {\bibfnamefont {F.}~\bibnamefont {Mila}}, \ and\ \bibinfo {author}
  {\bibfnamefont {A.}~\bibnamefont {Honecker}},\ }\href {\doibase
  10.1103/PhysRevResearch.1.033038} {\bibfield  {journal} {\bibinfo  {journal}
  {Phys. Rev. Research}\ }\textbf {\bibinfo {volume} {1}},\ \bibinfo {pages}
  {033038} (\bibinfo {year} {2019})}\BibitemShut {NoStop}%
\bibitem [{\citenamefont {Iitaka}\ and\ \citenamefont
  {Ebisuzaki}(2004)}]{Iitaka2004}%
  \BibitemOpen
  \bibfield  {author} {\bibinfo {author} {\bibfnamefont {T.}~\bibnamefont
  {Iitaka}}\ and\ \bibinfo {author} {\bibfnamefont {T.}~\bibnamefont
  {Ebisuzaki}},\ }\href {\doibase 10.1103/PhysRevE.69.057701} {\bibfield
  {journal} {\bibinfo  {journal} {Phys. Rev. E}\ }\textbf {\bibinfo {volume}
  {69}},\ \bibinfo {pages} {057701} (\bibinfo {year} {2004})}\BibitemShut
  {NoStop}%
\bibitem [{\citenamefont {Page}(1993)}]{Page1993}%
  \BibitemOpen
  \bibfield  {author} {\bibinfo {author} {\bibfnamefont {D.~N.}\ \bibnamefont
  {Page}},\ }\href {\doibase 10.1103/PhysRevLett.71.1291} {\bibfield  {journal}
  {\bibinfo  {journal} {Phys. Rev. Lett.}\ }\textbf {\bibinfo {volume} {71}},\
  \bibinfo {pages} {1291} (\bibinfo {year} {1993})}\BibitemShut {NoStop}%
\bibitem [{\citenamefont {Vidmar}\ and\ \citenamefont
  {Rigol}(2017)}]{Vidmar2017}%
  \BibitemOpen
  \bibfield  {author} {\bibinfo {author} {\bibfnamefont {L.}~\bibnamefont
  {Vidmar}}\ and\ \bibinfo {author} {\bibfnamefont {M.}~\bibnamefont {Rigol}},\
  }\href {\doibase 10.1103/PhysRevLett.119.220603} {\bibfield  {journal}
  {\bibinfo  {journal} {Phys. Rev. Lett.}\ }\textbf {\bibinfo {volume} {119}},\
  \bibinfo {pages} {220603} (\bibinfo {year} {2017})}\BibitemShut {NoStop}%
\bibitem [{\citenamefont {Schnack}\ \emph {et~al.}(2020)\citenamefont
  {Schnack}, \citenamefont {Richter},\ and\ \citenamefont
  {Steinigeweg}}]{Schnack2020}%
  \BibitemOpen
  \bibfield  {author} {\bibinfo {author} {\bibfnamefont {J.}~\bibnamefont
  {Schnack}}, \bibinfo {author} {\bibfnamefont {J.}~\bibnamefont {Richter}}, \
  and\ \bibinfo {author} {\bibfnamefont {R.}~\bibnamefont {Steinigeweg}},\
  }\href {\doibase 10.1103/PhysRevResearch.2.013186} {\bibfield  {journal}
  {\bibinfo  {journal} {Phys. Rev. Research}\ }\textbf {\bibinfo {volume}
  {2}},\ \bibinfo {pages} {013186} (\bibinfo {year} {2020})}\BibitemShut
  {NoStop}%
\bibitem [{\citenamefont {Monnai}\ and\ \citenamefont
  {Sugita}(2014)}]{Monnai2014}%
  \BibitemOpen
  \bibfield  {author} {\bibinfo {author} {\bibfnamefont {T.}~\bibnamefont
  {Monnai}}\ and\ \bibinfo {author} {\bibfnamefont {A.}~\bibnamefont
  {Sugita}},\ }\href {\doibase 10.7566/JPSJ.83.094001} {\bibfield  {journal}
  {\bibinfo  {journal} {J. Phys. Soc. Jpn.}\ }\textbf {\bibinfo {volume}
  {83}},\ \bibinfo {pages} {094001} (\bibinfo {year} {2014})}\BibitemShut
  {NoStop}%
\bibitem [{\citenamefont {Endo}\ \emph {et~al.}(2018)\citenamefont {Endo},
  \citenamefont {Hotta},\ and\ \citenamefont {Shimizu}}]{Endo2018}%
  \BibitemOpen
  \bibfield  {author} {\bibinfo {author} {\bibfnamefont {H.}~\bibnamefont
  {Endo}}, \bibinfo {author} {\bibfnamefont {C.}~\bibnamefont {Hotta}}, \ and\
  \bibinfo {author} {\bibfnamefont {A.}~\bibnamefont {Shimizu}},\ }\href
  {\doibase 10.1103/PhysRevLett.121.220601} {\bibfield  {journal} {\bibinfo
  {journal} {Phys. Rev. Lett.}\ }\textbf {\bibinfo {volume} {121}},\ \bibinfo
  {pages} {220601} (\bibinfo {year} {2018})}\BibitemShut {NoStop}%
\bibitem [{\citenamefont {Richter}\ \emph
  {et~al.}(2019{\natexlab{a}})\citenamefont {Richter}, \citenamefont {Lamann},
  \citenamefont {Bartsch}, \citenamefont {Steinigeweg},\ and\ \citenamefont
  {Gemmer}}]{Richter2019d}%
  \BibitemOpen
  \bibfield  {author} {\bibinfo {author} {\bibfnamefont {J.}~\bibnamefont
  {Richter}}, \bibinfo {author} {\bibfnamefont {M.~H.}\ \bibnamefont {Lamann}},
  \bibinfo {author} {\bibfnamefont {C.}~\bibnamefont {Bartsch}}, \bibinfo
  {author} {\bibfnamefont {R.}~\bibnamefont {Steinigeweg}}, \ and\ \bibinfo
  {author} {\bibfnamefont {J.}~\bibnamefont {Gemmer}},\ }\href {\doibase
  10.1103/PhysRevE.100.032124} {\bibfield  {journal} {\bibinfo  {journal}
  {Phys. Rev. E}\ }\textbf {\bibinfo {volume} {100}},\ \bibinfo {pages}
  {032124} (\bibinfo {year} {2019}{\natexlab{a}})}\BibitemShut {NoStop}%
\bibitem [{\citenamefont {Deutsch}(1991)}]{Deutsch1991}%
  \BibitemOpen
  \bibfield  {author} {\bibinfo {author} {\bibfnamefont {J.~M.}\ \bibnamefont
  {Deutsch}},\ }\href {\doibase 10.1103/PhysRevA.43.2046} {\bibfield  {journal}
  {\bibinfo  {journal} {Phys. Rev. A}\ }\textbf {\bibinfo {volume} {43}},\
  \bibinfo {pages} {2046} (\bibinfo {year} {1991})}\BibitemShut {NoStop}%
\bibitem [{\citenamefont {Srednicki}(1994)}]{Srednicki1994}%
  \BibitemOpen
  \bibfield  {author} {\bibinfo {author} {\bibfnamefont {M.}~\bibnamefont
  {Srednicki}},\ }\href {\doibase 10.1103/PhysRevE.50.888} {\bibfield
  {journal} {\bibinfo  {journal} {Phys. Rev. E}\ }\textbf {\bibinfo {volume}
  {50}},\ \bibinfo {pages} {888} (\bibinfo {year} {1994})}\BibitemShut
  {NoStop}%
\bibitem [{\citenamefont {Rigol}\ \emph {et~al.}(2008)\citenamefont {Rigol},
  \citenamefont {Dunjko},\ and\ \citenamefont {Olshanii}}]{Rigol2008}%
  \BibitemOpen
  \bibfield  {author} {\bibinfo {author} {\bibfnamefont {M.}~\bibnamefont
  {Rigol}}, \bibinfo {author} {\bibfnamefont {V.}~\bibnamefont {Dunjko}}, \
  and\ \bibinfo {author} {\bibfnamefont {M.}~\bibnamefont {Olshanii}},\ }\href
  {\doibase 10.1038/nature06838} {\bibfield  {journal} {\bibinfo  {journal}
  {Nature}\ }\textbf {\bibinfo {volume} {452}},\ \bibinfo {pages} {854}
  (\bibinfo {year} {2008})}\BibitemShut {NoStop}%
\bibitem [{\citenamefont {Santos}\ and\ \citenamefont
  {Rigol}(2010)}]{Santos2010}%
  \BibitemOpen
  \bibfield  {author} {\bibinfo {author} {\bibfnamefont {L.~F.}\ \bibnamefont
  {Santos}}\ and\ \bibinfo {author} {\bibfnamefont {M.}~\bibnamefont {Rigol}},\
  }\href {\doibase 10.1103/PhysRevE.82.031130} {\bibfield  {journal} {\bibinfo
  {journal} {Phys. Rev. E}\ }\textbf {\bibinfo {volume} {82}},\ \bibinfo
  {pages} {031130} (\bibinfo {year} {2010})}\BibitemShut {NoStop}%
\bibitem [{\citenamefont {Steinigeweg}\ \emph {et~al.}(2013)\citenamefont
  {Steinigeweg}, \citenamefont {Herbrych},\ and\ \citenamefont
  {Prelov{\v{s}}ek}}]{Steinigeweg2013}%
  \BibitemOpen
  \bibfield  {author} {\bibinfo {author} {\bibfnamefont {R.}~\bibnamefont
  {Steinigeweg}}, \bibinfo {author} {\bibfnamefont {J.}~\bibnamefont
  {Herbrych}}, \ and\ \bibinfo {author} {\bibfnamefont {P.}~\bibnamefont
  {Prelov{\v{s}}ek}},\ }\href {\doibase 10.1103/PhysRevE.87.012118} {\bibfield
  {journal} {\bibinfo  {journal} {Phys. Rev. E}\ }\textbf {\bibinfo {volume}
  {87}},\ \bibinfo {pages} {012118} (\bibinfo {year} {2013})}\BibitemShut
  {NoStop}%
\bibitem [{\citenamefont {Beugeling}\ \emph {et~al.}(2014)\citenamefont
  {Beugeling}, \citenamefont {Moessner},\ and\ \citenamefont
  {Haque}}]{Beugeling2014}%
  \BibitemOpen
  \bibfield  {author} {\bibinfo {author} {\bibfnamefont {W.}~\bibnamefont
  {Beugeling}}, \bibinfo {author} {\bibfnamefont {R.}~\bibnamefont {Moessner}},
  \ and\ \bibinfo {author} {\bibfnamefont {M.}~\bibnamefont {Haque}},\ }\href
  {\doibase 10.1103/PhysRevE.89.042112} {\bibfield  {journal} {\bibinfo
  {journal} {Phys. Rev. E}\ }\textbf {\bibinfo {volume} {89}},\ \bibinfo
  {pages} {042112} (\bibinfo {year} {2014})}\BibitemShut {NoStop}%
\bibitem [{\citenamefont {Kim}\ \emph {et~al.}(2014)\citenamefont {Kim},
  \citenamefont {Ikeda},\ and\ \citenamefont {Huse}}]{Kim2014}%
  \BibitemOpen
  \bibfield  {author} {\bibinfo {author} {\bibfnamefont {H.}~\bibnamefont
  {Kim}}, \bibinfo {author} {\bibfnamefont {T.~N.}\ \bibnamefont {Ikeda}}, \
  and\ \bibinfo {author} {\bibfnamefont {D.~A.}\ \bibnamefont {Huse}},\ }\href
  {\doibase 10.1103/PhysRevE.90.052105} {\bibfield  {journal} {\bibinfo
  {journal} {Phys. Rev. E}\ }\textbf {\bibinfo {volume} {90}},\ \bibinfo
  {pages} {052105} (\bibinfo {year} {2014})}\BibitemShut {NoStop}%
\bibitem [{\citenamefont {Nandkishore}\ and\ \citenamefont
  {Huse}(2015)}]{Nandkishore2015}%
  \BibitemOpen
  \bibfield  {author} {\bibinfo {author} {\bibfnamefont {R.}~\bibnamefont
  {Nandkishore}}\ and\ \bibinfo {author} {\bibfnamefont {D.~A.}\ \bibnamefont
  {Huse}},\ }\href {\doibase 10.1146/annurev-conmatphys-031214-014726}
  {\bibfield  {journal} {\bibinfo  {journal} {Annu. Rev. Condens. Matter
  Phys.}\ }\textbf {\bibinfo {volume} {6}},\ \bibinfo {pages} {15} (\bibinfo
  {year} {2015})}\BibitemShut {NoStop}%
\bibitem [{\citenamefont {Mondaini}\ and\ \citenamefont
  {Rigol}(2017)}]{Mondaini2017}%
  \BibitemOpen
  \bibfield  {author} {\bibinfo {author} {\bibfnamefont {R.}~\bibnamefont
  {Mondaini}}\ and\ \bibinfo {author} {\bibfnamefont {M.}~\bibnamefont
  {Rigol}},\ }\href {\doibase 10.1103/PhysRevE.96.012157} {\bibfield  {journal}
  {\bibinfo  {journal} {Phys. Rev. E}\ }\textbf {\bibinfo {volume} {96}},\
  \bibinfo {pages} {012157} (\bibinfo {year} {2017})}\BibitemShut {NoStop}%
\bibitem [{\citenamefont {Jansen}\ \emph {et~al.}(2019)\citenamefont {Jansen},
  \citenamefont {Stolpp}, \citenamefont {Vidmar},\ and\ \citenamefont
  {Heidrich-Meisner}}]{Jansen2019}%
  \BibitemOpen
  \bibfield  {author} {\bibinfo {author} {\bibfnamefont {D.}~\bibnamefont
  {Jansen}}, \bibinfo {author} {\bibfnamefont {J.}~\bibnamefont {Stolpp}},
  \bibinfo {author} {\bibfnamefont {L.}~\bibnamefont {Vidmar}}, \ and\ \bibinfo
  {author} {\bibfnamefont {F.}~\bibnamefont {Heidrich-Meisner}},\ }\href
  {\doibase 10.1103/PhysRevB.99.155130} {\bibfield  {journal} {\bibinfo
  {journal} {Phys. Rev. B}\ }\textbf {\bibinfo {volume} {99}},\ \bibinfo
  {pages} {155130} (\bibinfo {year} {2019})}\BibitemShut {NoStop}%
\bibitem [{\citenamefont {Khaymovich}\ \emph {et~al.}(2019)\citenamefont
  {Khaymovich}, \citenamefont {Haque},\ and\ \citenamefont
  {McClarty}}]{Khaymovich2019}%
  \BibitemOpen
  \bibfield  {author} {\bibinfo {author} {\bibfnamefont {I.~M.}\ \bibnamefont
  {Khaymovich}}, \bibinfo {author} {\bibfnamefont {M.}~\bibnamefont {Haque}}, \
  and\ \bibinfo {author} {\bibfnamefont {P.~A.}\ \bibnamefont {McClarty}},\
  }\href {\doibase 10.1103/PhysRevLett.122.070601} {\bibfield  {journal}
  {\bibinfo  {journal} {Phys. Rev. Lett.}\ }\textbf {\bibinfo {volume} {122}},\
  \bibinfo {pages} {070601} (\bibinfo {year} {2019})}\BibitemShut {NoStop}%
\bibitem [{\citenamefont {Steinigeweg}\ \emph
  {et~al.}(2016{\natexlab{a}})\citenamefont {Steinigeweg}, \citenamefont
  {Herbrych}, \citenamefont {Pollmann},\ and\ \citenamefont
  {Brenig}}]{Steinigeweg2016a}%
  \BibitemOpen
  \bibfield  {author} {\bibinfo {author} {\bibfnamefont {R.}~\bibnamefont
  {Steinigeweg}}, \bibinfo {author} {\bibfnamefont {J.}~\bibnamefont
  {Herbrych}}, \bibinfo {author} {\bibfnamefont {F.}~\bibnamefont {Pollmann}},
  \ and\ \bibinfo {author} {\bibfnamefont {W.}~\bibnamefont {Brenig}},\ }\href
  {\doibase 10.1103/PhysRevB.94.180401} {\bibfield  {journal} {\bibinfo
  {journal} {Phys. Rev. B}\ }\textbf {\bibinfo {volume} {94}},\ \bibinfo
  {pages} {180401} (\bibinfo {year} {2016}{\natexlab{a}})}\BibitemShut
  {NoStop}%
\bibitem [{\citenamefont {Steinigeweg}\ \emph
  {et~al.}(2017{\natexlab{a}})\citenamefont {Steinigeweg}, \citenamefont {Jin},
  \citenamefont {Schmidtke}, \citenamefont {{De Raedt}}, \citenamefont
  {Michielsen},\ and\ \citenamefont {Gemmer}}]{Steinigeweg2017a}%
  \BibitemOpen
  \bibfield  {author} {\bibinfo {author} {\bibfnamefont {R.}~\bibnamefont
  {Steinigeweg}}, \bibinfo {author} {\bibfnamefont {F.}~\bibnamefont {Jin}},
  \bibinfo {author} {\bibfnamefont {D.}~\bibnamefont {Schmidtke}}, \bibinfo
  {author} {\bibfnamefont {H.}~\bibnamefont {{De Raedt}}}, \bibinfo {author}
  {\bibfnamefont {K.}~\bibnamefont {Michielsen}}, \ and\ \bibinfo {author}
  {\bibfnamefont {J.}~\bibnamefont {Gemmer}},\ }\href {\doibase
  10.1103/PhysRevB.95.035155} {\bibfield  {journal} {\bibinfo  {journal} {Phys.
  Rev. B}\ }\textbf {\bibinfo {volume} {95}},\ \bibinfo {pages} {035155}
  (\bibinfo {year} {2017}{\natexlab{a}})}\BibitemShut {NoStop}%
\bibitem [{\citenamefont {Richter}\ \emph
  {et~al.}(2018{\natexlab{a}})\citenamefont {Richter}, \citenamefont
  {Herbrych},\ and\ \citenamefont {Steinigeweg}}]{Richter2018}%
  \BibitemOpen
  \bibfield  {author} {\bibinfo {author} {\bibfnamefont {J.}~\bibnamefont
  {Richter}}, \bibinfo {author} {\bibfnamefont {J.}~\bibnamefont {Herbrych}}, \
  and\ \bibinfo {author} {\bibfnamefont {R.}~\bibnamefont {Steinigeweg}},\
  }\href {\doibase 10.1103/PhysRevB.98.134302} {\bibfield  {journal} {\bibinfo
  {journal} {Phys. Rev. B}\ }\textbf {\bibinfo {volume} {98}},\ \bibinfo
  {pages} {134302} (\bibinfo {year} {2018}{\natexlab{a}})}\BibitemShut
  {NoStop}%
\bibitem [{\citenamefont {Richter}\ \emph
  {et~al.}(2019{\natexlab{b}})\citenamefont {Richter}, \citenamefont {Casper},
  \citenamefont {Brenig},\ and\ \citenamefont {Steinigeweg}}]{Richter2019g}%
  \BibitemOpen
  \bibfield  {author} {\bibinfo {author} {\bibfnamefont {J.}~\bibnamefont
  {Richter}}, \bibinfo {author} {\bibfnamefont {N.}~\bibnamefont {Casper}},
  \bibinfo {author} {\bibfnamefont {W.}~\bibnamefont {Brenig}}, \ and\ \bibinfo
  {author} {\bibfnamefont {R.}~\bibnamefont {Steinigeweg}},\ }\href {\doibase
  10.1103/PhysRevB.100.144423} {\bibfield  {journal} {\bibinfo  {journal}
  {Phys. Rev. B}\ }\textbf {\bibinfo {volume} {100}},\ \bibinfo {pages}
  {144423} (\bibinfo {year} {2019}{\natexlab{b}})}\BibitemShut {NoStop}%
\bibitem [{\citenamefont {Steinigeweg}\ \emph
  {et~al.}(2014{\natexlab{b}})\citenamefont {Steinigeweg}, \citenamefont
  {Khodja}, \citenamefont {Niemeyer}, \citenamefont {Gogolin},\ and\
  \citenamefont {Gemmer}}]{Steinigeweg2014a}%
  \BibitemOpen
  \bibfield  {author} {\bibinfo {author} {\bibfnamefont {R.}~\bibnamefont
  {Steinigeweg}}, \bibinfo {author} {\bibfnamefont {A.}~\bibnamefont {Khodja}},
  \bibinfo {author} {\bibfnamefont {H.}~\bibnamefont {Niemeyer}}, \bibinfo
  {author} {\bibfnamefont {C.}~\bibnamefont {Gogolin}}, \ and\ \bibinfo
  {author} {\bibfnamefont {J.}~\bibnamefont {Gemmer}},\ }\href {\doibase
  10.1103/PhysRevLett.112.130403} {\bibfield  {journal} {\bibinfo  {journal}
  {Phys. Rev. Lett.}\ }\textbf {\bibinfo {volume} {112}},\ \bibinfo {pages}
  {130403} (\bibinfo {year} {2014}{\natexlab{b}})}\BibitemShut {NoStop}%
\bibitem [{\citenamefont {Heitmann}\ and\ \citenamefont
  {Schnack}(2019)}]{Heitmann2019}%
  \BibitemOpen
  \bibfield  {author} {\bibinfo {author} {\bibfnamefont {T.}~\bibnamefont
  {Heitmann}}\ and\ \bibinfo {author} {\bibfnamefont {J.}~\bibnamefont
  {Schnack}},\ }\href {\doibase 10.1103/PhysRevB.99.134405} {\bibfield
  {journal} {\bibinfo  {journal} {Phys. Rev. B}\ }\textbf {\bibinfo {volume}
  {99}},\ \bibinfo {pages} {134405} (\bibinfo {year} {2019})}\BibitemShut
  {NoStop}%
\bibitem [{\citenamefont {Hughston}\ \emph {et~al.}(1993)\citenamefont
  {Hughston}, \citenamefont {Jozsa},\ and\ \citenamefont
  {Wootters}}]{Hughston1993}%
  \BibitemOpen
  \bibfield  {author} {\bibinfo {author} {\bibfnamefont {L.~P.}\ \bibnamefont
  {Hughston}}, \bibinfo {author} {\bibfnamefont {R.}~\bibnamefont {Jozsa}}, \
  and\ \bibinfo {author} {\bibfnamefont {W.~K.}\ \bibnamefont {Wootters}},\
  }\href {\doibase 10.1016/0375-9601(93)90880-9} {\bibfield  {journal}
  {\bibinfo  {journal} {Phys. Lett. A}\ }\textbf {\bibinfo {volume} {183}},\
  \bibinfo {pages} {14} (\bibinfo {year} {1993})}\BibitemShut {NoStop}%
\bibitem [{\citenamefont {{De Raedt}}\ and\ \citenamefont
  {Michielsen}(2006)}]{DeRaedt2006}%
  \BibitemOpen
  \bibfield  {author} {\bibinfo {author} {\bibfnamefont {H.}~\bibnamefont {{De
  Raedt}}}\ and\ \bibinfo {author} {\bibfnamefont {K.}~\bibnamefont
  {Michielsen}},\ }\href
  {http://hdl.handle.net/11370/00a4be1c-c9dd-4f34-92e5-2248b5f2a7d7
  http://arxiv.org/abs/quant-ph/0406210} {\bibfield  {journal} {\bibinfo
  {journal} {Host Publ.}\ } (\bibinfo {year} {2006})}\BibitemShut {NoStop}%
\bibitem [{\citenamefont {Nauts}\ and\ \citenamefont
  {Wyatt}(1983)}]{Nauts1983}%
  \BibitemOpen
  \bibfield  {author} {\bibinfo {author} {\bibfnamefont {A.}~\bibnamefont
  {Nauts}}\ and\ \bibinfo {author} {\bibfnamefont {R.~E.}\ \bibnamefont
  {Wyatt}},\ }\href {\doibase 10.1103/PhysRevLett.51.2238} {\bibfield
  {journal} {\bibinfo  {journal} {Phys. Rev. Lett.}\ }\textbf {\bibinfo
  {volume} {51}},\ \bibinfo {pages} {2238} (\bibinfo {year}
  {1983})}\BibitemShut {NoStop}%
\bibitem [{\citenamefont {Tal‐Ezer}\ and\ \citenamefont
  {Kosloff}(1984)}]{TalEzer1984}%
  \BibitemOpen
  \bibfield  {author} {\bibinfo {author} {\bibfnamefont {H.}~\bibnamefont
  {Tal‐Ezer}}\ and\ \bibinfo {author} {\bibfnamefont {R.}~\bibnamefont
  {Kosloff}},\ }\href {\doibase 10.1063/1.448136} {\bibfield  {journal}
  {\bibinfo  {journal} {J. Chem. Phys.}\ }\textbf {\bibinfo {volume} {81}},\
  \bibinfo {pages} {3967} (\bibinfo {year} {1984})}\BibitemShut {NoStop}%
\bibitem [{\citenamefont {Kosloff}(1994)}]{Kosloff1994}%
  \BibitemOpen
  \bibfield  {author} {\bibinfo {author} {\bibfnamefont {R.}~\bibnamefont
  {Kosloff}},\ }\href {\doibase 10.1146/annurev.physchem.45.1.145} {\bibfield
  {journal} {\bibinfo  {journal} {Annu. Rev. Phys. Chem.}\ }\textbf {\bibinfo
  {volume} {45}},\ \bibinfo {pages} {145} (\bibinfo {year} {1994})}\BibitemShut
  {NoStop}%
\bibitem [{\citenamefont {Dobrovitski}\ and\ \citenamefont {{De
  Raedt}}(2003)}]{Dobrovitski2003}%
  \BibitemOpen
  \bibfield  {author} {\bibinfo {author} {\bibfnamefont {V.~V.}\ \bibnamefont
  {Dobrovitski}}\ and\ \bibinfo {author} {\bibfnamefont {H.}~\bibnamefont {{De
  Raedt}}},\ }\href {\doibase 10.1103/PhysRevE.67.056702} {\bibfield  {journal}
  {\bibinfo  {journal} {Phys. Rev. E}\ }\textbf {\bibinfo {volume} {67}},\
  \bibinfo {pages} {056702} (\bibinfo {year} {2003})}\BibitemShut {NoStop}%
\bibitem [{\citenamefont {Wei{\ss}e}\ \emph {et~al.}(2006)\citenamefont
  {Wei{\ss}e}, \citenamefont {Wellein}, \citenamefont {Alvermann},\ and\
  \citenamefont {Fehske}}]{Weisse2006}%
  \BibitemOpen
  \bibfield  {author} {\bibinfo {author} {\bibfnamefont {A.}~\bibnamefont
  {Wei{\ss}e}}, \bibinfo {author} {\bibfnamefont {G.}~\bibnamefont {Wellein}},
  \bibinfo {author} {\bibfnamefont {A.}~\bibnamefont {Alvermann}}, \ and\
  \bibinfo {author} {\bibfnamefont {H.}~\bibnamefont {Fehske}},\ }\href
  {\doibase 10.1103/RevModPhys.78.275} {\bibfield  {journal} {\bibinfo
  {journal} {Rev. Mod. Phys.}\ }\textbf {\bibinfo {volume} {78}},\ \bibinfo
  {pages} {275} (\bibinfo {year} {2006})}\BibitemShut {NoStop}%
\bibitem [{\citenamefont {Kosloff}(2019)}]{Kosloff2019}%
  \BibitemOpen
  \bibfield  {author} {\bibinfo {author} {\bibfnamefont {R.}~\bibnamefont
  {Kosloff}},\ }\href {\doibase 10.1063/1.5096173} {\bibfield  {journal}
  {\bibinfo  {journal} {J. Chem. Phys.}\ }\textbf {\bibinfo {volume} {150}},\
  \bibinfo {pages} {204105} (\bibinfo {year} {2019})}\BibitemShut {NoStop}%
\bibitem [{\citenamefont {Schiulaz}\ \emph {et~al.}(2019)\citenamefont
  {Schiulaz}, \citenamefont {Torres-Herrera},\ and\ \citenamefont
  {Santos}}]{Schiulaz2019}%
  \BibitemOpen
  \bibfield  {author} {\bibinfo {author} {\bibfnamefont {M.}~\bibnamefont
  {Schiulaz}}, \bibinfo {author} {\bibfnamefont {E.~J.}\ \bibnamefont
  {Torres-Herrera}}, \ and\ \bibinfo {author} {\bibfnamefont {L.~F.}\
  \bibnamefont {Santos}},\ }\href {\doibase 10.1103/PhysRevB.99.174313}
  {\bibfield  {journal} {\bibinfo  {journal} {Phys. Rev. B}\ }\textbf {\bibinfo
  {volume} {99}},\ \bibinfo {pages} {174313} (\bibinfo {year}
  {2019})}\BibitemShut {NoStop}%
\bibitem [{\citenamefont {Richter}\ \emph
  {et~al.}(2018{\natexlab{b}})\citenamefont {Richter}, \citenamefont {Jin},
  \citenamefont {{De Raedt}}, \citenamefont {Michielsen}, \citenamefont
  {Gemmer},\ and\ \citenamefont {Steinigeweg}}]{Richter2018a}%
  \BibitemOpen
  \bibfield  {author} {\bibinfo {author} {\bibfnamefont {J.}~\bibnamefont
  {Richter}}, \bibinfo {author} {\bibfnamefont {F.}~\bibnamefont {Jin}},
  \bibinfo {author} {\bibfnamefont {H.}~\bibnamefont {{De Raedt}}}, \bibinfo
  {author} {\bibfnamefont {K.}~\bibnamefont {Michielsen}}, \bibinfo {author}
  {\bibfnamefont {J.}~\bibnamefont {Gemmer}}, \ and\ \bibinfo {author}
  {\bibfnamefont {R.}~\bibnamefont {Steinigeweg}},\ }\href {\doibase
  10.1103/PhysRevB.97.174430} {\bibfield  {journal} {\bibinfo  {journal} {Phys.
  Rev. B}\ }\textbf {\bibinfo {volume} {97}},\ \bibinfo {pages} {174430}
  (\bibinfo {year} {2018}{\natexlab{b}})}\BibitemShut {NoStop}%
\bibitem [{\citenamefont {Kennes}\ \emph {et~al.}(2020)\citenamefont {Kennes},
  \citenamefont {Karrasch},\ and\ \citenamefont {Millis}}]{Kennes2020}%
  \BibitemOpen
  \bibfield  {author} {\bibinfo {author} {\bibfnamefont {D.~M.}\ \bibnamefont
  {Kennes}}, \bibinfo {author} {\bibfnamefont {C.}~\bibnamefont {Karrasch}}, \
  and\ \bibinfo {author} {\bibfnamefont {A.~J.}\ \bibnamefont {Millis}},\
  }\href {\doibase 10.1103/PhysRevB.101.081106} {\bibfield  {journal} {\bibinfo
   {journal} {Phys. Rev. B}\ }\textbf {\bibinfo {volume} {101}},\ \bibinfo
  {pages} {081106} (\bibinfo {year} {2020})}\BibitemShut {NoStop}%
\bibitem [{\citenamefont {Heidrich-Meisner}\ \emph {et~al.}(2007)\citenamefont
  {Heidrich-Meisner}, \citenamefont {Honecker},\ and\ \citenamefont
  {Brenig}}]{Heidrich-Meisner2007}%
  \BibitemOpen
  \bibfield  {author} {\bibinfo {author} {\bibfnamefont {F.}~\bibnamefont
  {Heidrich-Meisner}}, \bibinfo {author} {\bibfnamefont {A.}~\bibnamefont
  {Honecker}}, \ and\ \bibinfo {author} {\bibfnamefont {W.}~\bibnamefont
  {Brenig}},\ }\href {\doibase 10.1140/epjst/e2007-00369-2} {\bibfield
  {journal} {\bibinfo  {journal} {Eur. Phys. J. Spec. Top.}\ }\textbf {\bibinfo
  {volume} {151}},\ \bibinfo {pages} {135} (\bibinfo {year}
  {2007})}\BibitemShut {NoStop}%
\bibitem [{\citenamefont {Steinigeweg}\ \emph {et~al.}(2015)\citenamefont
  {Steinigeweg}, \citenamefont {Gemmer},\ and\ \citenamefont
  {Brenig}}]{Steinigeweg2015}%
  \BibitemOpen
  \bibfield  {author} {\bibinfo {author} {\bibfnamefont {R.}~\bibnamefont
  {Steinigeweg}}, \bibinfo {author} {\bibfnamefont {J.}~\bibnamefont {Gemmer}},
  \ and\ \bibinfo {author} {\bibfnamefont {W.}~\bibnamefont {Brenig}},\ }\href
  {\doibase 10.1103/PhysRevB.91.104404} {\bibfield  {journal} {\bibinfo
  {journal} {Phys. Rev. B}\ }\textbf {\bibinfo {volume} {91}},\ \bibinfo
  {pages} {104404} (\bibinfo {year} {2015})}\BibitemShut {NoStop}%
\bibitem [{\citenamefont {Steinigeweg}\ \emph
  {et~al.}(2014{\natexlab{c}})\citenamefont {Steinigeweg}, \citenamefont
  {Heidrich-Meisner}, \citenamefont {Gemmer}, \citenamefont {Michielsen},\ and\
  \citenamefont {{De Raedt}}}]{Steinigeweg2014b}%
  \BibitemOpen
  \bibfield  {author} {\bibinfo {author} {\bibfnamefont {R.}~\bibnamefont
  {Steinigeweg}}, \bibinfo {author} {\bibfnamefont {F.}~\bibnamefont
  {Heidrich-Meisner}}, \bibinfo {author} {\bibfnamefont {J.}~\bibnamefont
  {Gemmer}}, \bibinfo {author} {\bibfnamefont {K.}~\bibnamefont {Michielsen}},
  \ and\ \bibinfo {author} {\bibfnamefont {H.}~\bibnamefont {{De Raedt}}},\
  }\href {\doibase 10.1103/PhysRevB.90.094417} {\bibfield  {journal} {\bibinfo
  {journal} {Phys. Rev. B}\ }\textbf {\bibinfo {volume} {90}},\ \bibinfo
  {pages} {094417} (\bibinfo {year} {2014}{\natexlab{c}})}\BibitemShut
  {NoStop}%
\bibitem [{\citenamefont {Steinigeweg}\ \emph
  {et~al.}(2016{\natexlab{b}})\citenamefont {Steinigeweg}, \citenamefont
  {Herbrych}, \citenamefont {Zotos},\ and\ \citenamefont
  {Brenig}}]{Steinigeweg2016}%
  \BibitemOpen
  \bibfield  {author} {\bibinfo {author} {\bibfnamefont {R.}~\bibnamefont
  {Steinigeweg}}, \bibinfo {author} {\bibfnamefont {J.}~\bibnamefont
  {Herbrych}}, \bibinfo {author} {\bibfnamefont {X.}~\bibnamefont {Zotos}}, \
  and\ \bibinfo {author} {\bibfnamefont {W.}~\bibnamefont {Brenig}},\ }\href
  {\doibase 10.1103/PhysRevLett.116.017202} {\bibfield  {journal} {\bibinfo
  {journal} {Phys. Rev. Lett.}\ }\textbf {\bibinfo {volume} {116}},\ \bibinfo
  {pages} {017202} (\bibinfo {year} {2016}{\natexlab{b}})}\BibitemShut
  {NoStop}%
\bibitem [{\citenamefont {Richter}\ \emph
  {et~al.}(2019{\natexlab{c}})\citenamefont {Richter}, \citenamefont {Jin},
  \citenamefont {Knipschild}, \citenamefont {Herbrych}, \citenamefont {{De
  Raedt}}, \citenamefont {Michielsen}, \citenamefont {Gemmer},\ and\
  \citenamefont {Steinigeweg}}]{Richter2019f}%
  \BibitemOpen
  \bibfield  {author} {\bibinfo {author} {\bibfnamefont {J.}~\bibnamefont
  {Richter}}, \bibinfo {author} {\bibfnamefont {F.}~\bibnamefont {Jin}},
  \bibinfo {author} {\bibfnamefont {L.}~\bibnamefont {Knipschild}}, \bibinfo
  {author} {\bibfnamefont {J.}~\bibnamefont {Herbrych}}, \bibinfo {author}
  {\bibfnamefont {H.}~\bibnamefont {{De Raedt}}}, \bibinfo {author}
  {\bibfnamefont {K.}~\bibnamefont {Michielsen}}, \bibinfo {author}
  {\bibfnamefont {J.}~\bibnamefont {Gemmer}}, \ and\ \bibinfo {author}
  {\bibfnamefont {R.}~\bibnamefont {Steinigeweg}},\ }\href {\doibase
  10.1103/PhysRevB.99.144422} {\bibfield  {journal} {\bibinfo  {journal} {Phys.
  Rev. B}\ }\textbf {\bibinfo {volume} {99}},\ \bibinfo {pages} {144422}
  (\bibinfo {year} {2019}{\natexlab{c}})}\BibitemShut {NoStop}%
\bibitem [{\citenamefont {Jin}\ \emph {et~al.}(2015)\citenamefont {Jin},
  \citenamefont {Steinigeweg}, \citenamefont {Heidrich-Meisner}, \citenamefont
  {Michielsen},\ and\ \citenamefont {{De Raedt}}}]{Jin2015}%
  \BibitemOpen
  \bibfield  {author} {\bibinfo {author} {\bibfnamefont {F.}~\bibnamefont
  {Jin}}, \bibinfo {author} {\bibfnamefont {R.}~\bibnamefont {Steinigeweg}},
  \bibinfo {author} {\bibfnamefont {F.}~\bibnamefont {Heidrich-Meisner}},
  \bibinfo {author} {\bibfnamefont {K.}~\bibnamefont {Michielsen}}, \ and\
  \bibinfo {author} {\bibfnamefont {H.}~\bibnamefont {{De Raedt}}},\ }\href
  {\doibase 10.1103/PhysRevB.92.205103} {\bibfield  {journal} {\bibinfo
  {journal} {Phys. Rev. B}\ }\textbf {\bibinfo {volume} {92}},\ \bibinfo
  {pages} {205103} (\bibinfo {year} {2015})}\BibitemShut {NoStop}%
\bibitem [{\citenamefont {Balz}\ \emph {et~al.}(2018)\citenamefont {Balz},
  \citenamefont {Richter}, \citenamefont {Gemmer}, \citenamefont
  {Steinigeweg},\ and\ \citenamefont {Reimann}}]{Balz2019}%
  \BibitemOpen
  \bibfield  {author} {\bibinfo {author} {\bibfnamefont {B.~N.}\ \bibnamefont
  {Balz}}, \bibinfo {author} {\bibfnamefont {J.}~\bibnamefont {Richter}},
  \bibinfo {author} {\bibfnamefont {J.}~\bibnamefont {Gemmer}}, \bibinfo
  {author} {\bibfnamefont {R.}~\bibnamefont {Steinigeweg}}, \ and\ \bibinfo
  {author} {\bibfnamefont {P.}~\bibnamefont {Reimann}},\ }in\ \href {\doibase
  10.1007/978-3-319-99046-0_17} {\emph {\bibinfo {booktitle} {Thermodynamics in
  the Quantum Regime}}}\ (\bibinfo  {publisher} {Springer, Cham},\ \bibinfo
  {year} {2018})\ pp.\ \bibinfo {pages} {413--433}\BibitemShut {NoStop}%
\bibitem [{\citenamefont {Steinigeweg}\ \emph
  {et~al.}(2017{\natexlab{b}})\citenamefont {Steinigeweg}, \citenamefont {Jin},
  \citenamefont {{De Raedt}}, \citenamefont {Michielsen},\ and\ \citenamefont
  {Gemmer}}]{Steinigeweg2017}%
  \BibitemOpen
  \bibfield  {author} {\bibinfo {author} {\bibfnamefont {R.}~\bibnamefont
  {Steinigeweg}}, \bibinfo {author} {\bibfnamefont {F.}~\bibnamefont {Jin}},
  \bibinfo {author} {\bibfnamefont {H.}~\bibnamefont {{De Raedt}}}, \bibinfo
  {author} {\bibfnamefont {K.}~\bibnamefont {Michielsen}}, \ and\ \bibinfo
  {author} {\bibfnamefont {J.}~\bibnamefont {Gemmer}},\ }\href {\doibase
  10.1103/PhysRevE.96.020105} {\bibfield  {journal} {\bibinfo  {journal} {Phys.
  Rev. E}\ }\textbf {\bibinfo {volume} {96}},\ \bibinfo {pages} {020105}
  (\bibinfo {year} {2017}{\natexlab{b}})}\BibitemShut {NoStop}%
\bibitem [{\citenamefont {Ljubotina}\ \emph {et~al.}(2017)\citenamefont
  {Ljubotina}, \citenamefont {{\v{Z}}nidari{\v{c}}},\ and\ \citenamefont
  {Prosen}}]{Ljubotina2017}%
  \BibitemOpen
  \bibfield  {author} {\bibinfo {author} {\bibfnamefont {M.}~\bibnamefont
  {Ljubotina}}, \bibinfo {author} {\bibfnamefont {M.}~\bibnamefont
  {{\v{Z}}nidari{\v{c}}}}, \ and\ \bibinfo {author} {\bibfnamefont
  {T.}~\bibnamefont {Prosen}},\ }\href {\doibase 10.1038/ncomms16117}
  {\bibfield  {journal} {\bibinfo  {journal} {Nat. Commun.}\ }\textbf {\bibinfo
  {volume} {8}},\ \bibinfo {pages} {16117} (\bibinfo {year}
  {2017})}\BibitemShut {NoStop}%
\bibitem [{\citenamefont {Richter}\ \emph {et~al.}(2020)\citenamefont
  {Richter}, \citenamefont {Schubert},\ and\ \citenamefont
  {Steinigeweg}}]{Richter2020}%
  \BibitemOpen
  \bibfield  {author} {\bibinfo {author} {\bibfnamefont {J.}~\bibnamefont
  {Richter}}, \bibinfo {author} {\bibfnamefont {D.}~\bibnamefont {Schubert}}, \
  and\ \bibinfo {author} {\bibfnamefont {R.}~\bibnamefont {Steinigeweg}},\
  }\href {\doibase 10.1103/PhysRevResearch.2.013130} {\bibfield  {journal}
  {\bibinfo  {journal} {Phys. Rev. Research}\ }\textbf {\bibinfo {volume}
  {2}},\ \bibinfo {pages} {013130} (\bibinfo {year} {2020})}\BibitemShut
  {NoStop}%
\bibitem [{\citenamefont {Mitra}(2018)}]{Mitra2018}%
  \BibitemOpen
  \bibfield  {author} {\bibinfo {author} {\bibfnamefont {A.}~\bibnamefont
  {Mitra}},\ }\href {\doibase 10.1146/annurev-conmatphys-031016-025451}
  {\bibfield  {journal} {\bibinfo  {journal} {Annu. Rev. Condens. Matter
  Phys.}\ }\textbf {\bibinfo {volume} {9}},\ \bibinfo {pages} {245} (\bibinfo
  {year} {2018})}\BibitemShut {NoStop}%
\bibitem [{\citenamefont {Bartsch}\ and\ \citenamefont
  {Gemmer}(2017)}]{Bartsch2017}%
  \BibitemOpen
  \bibfield  {author} {\bibinfo {author} {\bibfnamefont {C.}~\bibnamefont
  {Bartsch}}\ and\ \bibinfo {author} {\bibfnamefont {J.}~\bibnamefont
  {Gemmer}},\ }\href {\doibase 10.1209/0295-5075/118/10006} {\bibfield
  {journal} {\bibinfo  {journal} {EPL (Europhys. Lett.)}\ }\textbf {\bibinfo
  {volume} {118}},\ \bibinfo {pages} {10006} (\bibinfo {year}
  {2017})}\BibitemShut {NoStop}%
\bibitem [{\citenamefont {Richter}\ \emph
  {et~al.}(2019{\natexlab{d}})\citenamefont {Richter}, \citenamefont {Gemmer},\
  and\ \citenamefont {Steinigeweg}}]{Richter2019a}%
  \BibitemOpen
  \bibfield  {author} {\bibinfo {author} {\bibfnamefont {J.}~\bibnamefont
  {Richter}}, \bibinfo {author} {\bibfnamefont {J.}~\bibnamefont {Gemmer}}, \
  and\ \bibinfo {author} {\bibfnamefont {R.}~\bibnamefont {Steinigeweg}},\
  }\href {\doibase 10.1103/PhysRevE.99.050104} {\bibfield  {journal} {\bibinfo
  {journal} {Phys. Rev. E}\ }\textbf {\bibinfo {volume} {99}},\ \bibinfo
  {pages} {050104} (\bibinfo {year} {2019}{\natexlab{d}})}\BibitemShut
  {NoStop}%
\bibitem [{\citenamefont {Rigol}\ \emph {et~al.}(2006)\citenamefont {Rigol},
  \citenamefont {Bryant},\ and\ \citenamefont {Singh}}]{Rigol2006b}%
  \BibitemOpen
  \bibfield  {author} {\bibinfo {author} {\bibfnamefont {M.}~\bibnamefont
  {Rigol}}, \bibinfo {author} {\bibfnamefont {T.}~\bibnamefont {Bryant}}, \
  and\ \bibinfo {author} {\bibfnamefont {R.~R.~P.}\ \bibnamefont {Singh}},\
  }\href {\doibase 10.1103/PhysRevLett.97.187202} {\bibfield  {journal}
  {\bibinfo  {journal} {Phys. Rev. Lett.}\ }\textbf {\bibinfo {volume} {97}},\
  \bibinfo {pages} {187202} (\bibinfo {year} {2006})}\BibitemShut {NoStop}%
\bibitem [{\citenamefont {Tang}\ \emph {et~al.}(2013)\citenamefont {Tang},
  \citenamefont {Khatami},\ and\ \citenamefont {Rigol}}]{Tang2013}%
  \BibitemOpen
  \bibfield  {author} {\bibinfo {author} {\bibfnamefont {B.}~\bibnamefont
  {Tang}}, \bibinfo {author} {\bibfnamefont {E.}~\bibnamefont {Khatami}}, \
  and\ \bibinfo {author} {\bibfnamefont {M.}~\bibnamefont {Rigol}},\ }\href
  {\doibase 10.1016/j.cpc.2012.10.008} {\bibfield  {journal} {\bibinfo
  {journal} {Comput. Phys. Commun.}\ }\textbf {\bibinfo {volume} {184}},\
  \bibinfo {pages} {557} (\bibinfo {year} {2013})}\BibitemShut {NoStop}%
\bibitem [{\citenamefont {Richter}\ and\ \citenamefont
  {Steinigeweg}(2019{\natexlab{b}})}]{Richter2019b}%
  \BibitemOpen
  \bibfield  {author} {\bibinfo {author} {\bibfnamefont {J.}~\bibnamefont
  {Richter}}\ and\ \bibinfo {author} {\bibfnamefont {R.}~\bibnamefont
  {Steinigeweg}},\ }\href {\doibase 10.1103/PhysRevB.99.094419} {\bibfield
  {journal} {\bibinfo  {journal} {Phys. Rev. B}\ }\textbf {\bibinfo {volume}
  {99}},\ \bibinfo {pages} {094419} (\bibinfo {year}
  {2019}{\natexlab{b}})}\BibitemShut {NoStop}%
\bibitem [{\citenamefont {White}\ \emph {et~al.}()\citenamefont {White},
  \citenamefont {Sundar},\ and\ \citenamefont {Hazzard}}]{White2017}%
  \BibitemOpen
  \bibfield  {author} {\bibinfo {author} {\bibfnamefont {I.~G.}\ \bibnamefont
  {White}}, \bibinfo {author} {\bibfnamefont {B.}~\bibnamefont {Sundar}}, \
  and\ \bibinfo {author} {\bibfnamefont {K.~R.~A.}\ \bibnamefont {Hazzard}},\
  }\href {http://arxiv.org/abs/1710.07696} {\ }\Eprint
  {http://arxiv.org/abs/1710.07696} {arXiv:1710.07696} \BibitemShut {NoStop}%
\bibitem [{\citenamefont {Mallayya}\ and\ \citenamefont
  {Rigol}(2018)}]{Mallayya2018}%
  \BibitemOpen
  \bibfield  {author} {\bibinfo {author} {\bibfnamefont {K.}~\bibnamefont
  {Mallayya}}\ and\ \bibinfo {author} {\bibfnamefont {M.}~\bibnamefont
  {Rigol}},\ }\href {\doibase 10.1103/PhysRevLett.120.070603} {\bibfield
  {journal} {\bibinfo  {journal} {Phys. Rev. Lett.}\ }\textbf {\bibinfo
  {volume} {120}},\ \bibinfo {pages} {070603} (\bibinfo {year}
  {2018})}\BibitemShut {NoStop}%
\bibitem [{\citenamefont {Bhattaram}\ and\ \citenamefont
  {Khatami}(2019)}]{Bhattaram2019}%
  \BibitemOpen
  \bibfield  {author} {\bibinfo {author} {\bibfnamefont {K.}~\bibnamefont
  {Bhattaram}}\ and\ \bibinfo {author} {\bibfnamefont {E.}~\bibnamefont
  {Khatami}},\ }\href {\doibase 10.1103/PhysRevE.100.013305} {\bibfield
  {journal} {\bibinfo  {journal} {Phys. Rev. E}\ }\textbf {\bibinfo {volume}
  {100}},\ \bibinfo {pages} {013305} (\bibinfo {year} {2019})}\BibitemShut
  {NoStop}%
\bibitem [{\citenamefont {Richter}\ \emph {et~al.}()\citenamefont {Richter},
  \citenamefont {Jin}, \citenamefont {Knipschild}, \citenamefont {{De Raedt}},
  \citenamefont {Michielsen}, \citenamefont {Gemmer},\ and\ \citenamefont
  {Steinigeweg}}]{Richter2019e}%
  \BibitemOpen
  \bibfield  {author} {\bibinfo {author} {\bibfnamefont {J.}~\bibnamefont
  {Richter}}, \bibinfo {author} {\bibfnamefont {F.}~\bibnamefont {Jin}},
  \bibinfo {author} {\bibfnamefont {L.}~\bibnamefont {Knipschild}}, \bibinfo
  {author} {\bibfnamefont {H.}~\bibnamefont {{De Raedt}}}, \bibinfo {author}
  {\bibfnamefont {K.}~\bibnamefont {Michielsen}}, \bibinfo {author}
  {\bibfnamefont {J.}~\bibnamefont {Gemmer}}, \ and\ \bibinfo {author}
  {\bibfnamefont {R.}~\bibnamefont {Steinigeweg}},\ }\href
  {http://arxiv.org/abs/1906.09268} {\ }\Eprint
  {http://arxiv.org/abs/1906.09268} {arXiv:1906.09268} \BibitemShut {NoStop}%
\bibitem [{\citenamefont {Chaturvedi}\ and\ \citenamefont
  {Shibata}(1979)}]{Chaturvedi1979}%
  \BibitemOpen
  \bibfield  {author} {\bibinfo {author} {\bibfnamefont {S.}~\bibnamefont
  {Chaturvedi}}\ and\ \bibinfo {author} {\bibfnamefont {F.}~\bibnamefont
  {Shibata}},\ }\href {\doibase 10.1007/BF01319852} {\bibfield  {journal}
  {\bibinfo  {journal} {Z. Phys. B}\ }\textbf {\bibinfo {volume} {35}},\
  \bibinfo {pages} {297} (\bibinfo {year} {1979})}\BibitemShut {NoStop}%
\bibitem [{\citenamefont {Breuer}\ and\ \citenamefont
  {Petruccione}(2007)}]{Breuer2007}%
  \BibitemOpen
  \bibfield  {author} {\bibinfo {author} {\bibfnamefont {H.-P.}\ \bibnamefont
  {Breuer}}\ and\ \bibinfo {author} {\bibfnamefont {F.}~\bibnamefont
  {Petruccione}},\ }\href {\doibase 10.1093/acprof:oso/9780199213900.001.0001}
  {\emph {\bibinfo {title} {The Theory of Open Quantum Systems}}}\ (\bibinfo
  {publisher} {Oxford University Press},\ \bibinfo {year} {2007})\BibitemShut
  {NoStop}%
\bibitem [{\citenamefont {Steinigeweg}(2011)}]{Steinigeweg2011b}%
  \BibitemOpen
  \bibfield  {author} {\bibinfo {author} {\bibfnamefont {R.}~\bibnamefont
  {Steinigeweg}},\ }\href {\doibase 10.1103/PhysRevE.84.011136} {\bibfield
  {journal} {\bibinfo  {journal} {Phys. Rev. E}\ }\textbf {\bibinfo {volume}
  {84}},\ \bibinfo {pages} {011136} (\bibinfo {year} {2011})}\BibitemShut
  {NoStop}%
\bibitem [{\citenamefont {Luitz}\ and\ \citenamefont {{Bar
  Lev}}(2017)}]{Luitz2017}%
  \BibitemOpen
  \bibfield  {author} {\bibinfo {author} {\bibfnamefont {D.~J.}\ \bibnamefont
  {Luitz}}\ and\ \bibinfo {author} {\bibfnamefont {Y.}~\bibnamefont {{Bar
  Lev}}},\ }\href {\doibase 10.1103/PhysRevB.96.020406} {\bibfield  {journal}
  {\bibinfo  {journal} {Phys. Rev. B}\ }\textbf {\bibinfo {volume} {96}},\
  \bibinfo {pages} {020406} (\bibinfo {year} {2017})}\BibitemShut {NoStop}%
\bibitem [{\citenamefont {Maldacena}\ \emph {et~al.}(2016)\citenamefont
  {Maldacena}, \citenamefont {Shenker},\ and\ \citenamefont
  {Stanford}}]{Maldacena2016}%
  \BibitemOpen
  \bibfield  {author} {\bibinfo {author} {\bibfnamefont {J.}~\bibnamefont
  {Maldacena}}, \bibinfo {author} {\bibfnamefont {S.~H.}\ \bibnamefont
  {Shenker}}, \ and\ \bibinfo {author} {\bibfnamefont {D.}~\bibnamefont
  {Stanford}},\ }\href {\doibase 10.1007/JHEP08(2016)106} {\bibfield  {journal}
  {\bibinfo  {journal} {J. High Energ. Phys.}\ }\textbf {\bibinfo {volume}
  {2016}},\ \bibinfo {pages} {106} (\bibinfo {year} {2016})}\BibitemShut
  {NoStop}%
\end{thebibliography}%

\end{document}